\documentclass[authoryear,preprint,12pt]{elsarticle}
\usepackage[utf8]{inputenc}
\usepackage{graphicx} % Required for inserting images
\usepackage{natbib}
\usepackage{tikz}
\usepackage{url}
\usepackage{xspace}
\usepackage{listings}
\usepackage{hyperref}
\usepackage{amsmath}
\usepackage{booktabs}    
\usepackage{multirow}    
\usepackage{caption} 
\usepackage{amssymb}
\usepackage{fontawesome5}
\usepackage{caption}
\usepackage[pagewise]{lineno}
\hypersetup{
    colorlinks=true,
    linkcolor=blue,
    filecolor=magenta,      
    urlcolor=cyan,
    pdftitle={revision}
    }
\modulolinenumbers[1]

\setcitestyle{authoryear}
\setcitestyle{citesep={;}}
\setcitestyle{open={(}}
\setcitestyle{close={)}}

\usetikzlibrary{shapes.geometric, arrows.meta, positioning, backgrounds, fit, decorations.pathmorphing, calc, external}
\tikzstyle{process} = [rectangle, minimum width=3cm, minimum height=1cm, text centered, draw=black, fill=white!30]
\tikzstyle{block} = [rectangle, rounded corners, minimum width=4cm, minimum height=1.2cm,text centered, draw=black, fill=white, font=\large]
\tikzstyle{data} = [rectangle, minimum width=4cm, minimum height=1.2cm, text centered, draw=black, fill=white!10, font=\large, thick]
\tikzstyle{arrow} = [ultra thick, ->, >=stealth, rounded corners=5pt]
\tikzstyle{dashedarrow} = [ultra thick, dashed, ->, >=stealth, rounded corners=5pt]
\tikzstyle{circleplus} = [circle, minimum size=1.2cm, draw=black, fill=blue!10, thick]

\newcommand{\n}{\ensuremath{\widetilde{N}}\xspace}
\newcommand{\ntf}{\ensuremath{\widetilde{N}[t_k, f_i]}\xspace}

\DeclareMathOperator*{\argmin}{arg\,min}
\definecolor{codegreen}{rgb}{0,0.6,0}
\definecolor{codegray}{rgb}{0.5,0.5,0.5}
\definecolor{codepurple}{rgb}{0.58,0,0.82}
\definecolor{backcolour}{rgb}{0.95,0.95,0.92}

\lstdefinestyle{mystyle}{
    backgroundcolor=\color{backcolour},   
    commentstyle=\color{codegreen},
    keywordstyle=\color{magenta},
    numberstyle=\tiny\color{codegray},
    stringstyle=\color{codepurple},
    basicstyle=\ttfamily\footnotesize,
    breakatwhitespace=false,         
    breaklines=true,                 
    captionpos=b,                    
    keepspaces=true,                 
    numbers=left,                    
    numbersep=5pt,                  
    showspaces=false,                
    showstringspaces=false,
    showtabs=false,                  
    tabsize=2
}
\journal{Hearing Research}

\begin{document}
% \linenumbers

\begin{frontmatter}

\title{From Spikes to Speech: NeuroVoc -- A Biologically Plausible Vocoder Framework for Auditory Perception and Cochlear Implant Simulation}

\author{Jacob de Nobel}
\author[lumc]{Jeroen J. Briaire}
\author[liacs]{Thomas H.W. B\"ack}
\author[liacs]{Anna V. Kononova}
\author[lumc,cog,bio]{Johan H.M. Frijns}

\address[liacs]{Leiden Institute of Advanced Computer Science, Niels Bohrweg 1, Leiden, Netherlands}
\address[lumc]{Department of Otorhinolaryngology, Leiden University Medical Center, Albinusdreef 2, Leiden, Netherlands}
\address[cog]{Leiden Institute for Brain and Cognition, Wassenaarseweg 52, Leiden, Netherlands}
\address[bio]{Bioelectronics group, EEMCS, Delft University of Technology, Mekelweg 4, Delft, Netherlands}

\begin{keyword}
Hearing Loss, Cochlear Implants, Auditory Periphery, Neural Model, Vocoder, Auditory Perception, Auditory Nerve, Neural Decoding, Signal Processing, Phenomological Model



\end{keyword}

\makeatletter
\def\@title{From Spikes to Speech: NeuroVoc\\
\large A Biologically Plausible Vocoder Framework for Auditory Perception and Cochlear Implant Simulation}
\makeatother
\end{frontmatter}

\section{Introduction}
In Cochlear Implant (CI) research, vocoders are often used as simulators to mimic how sound is processed and heard through a CI \citep{cychosz2024vocode}. Traditionally, the vocoder, which is a conjugation of the words \emph{voice} and \emph{encoder}, is a signal processing method used to break down and reconstruct speech material for efficient telecommunication \citep{dudley1939remaking}. Specifically, this so-called channel vocoder works by extracting the temporal envelope of an audio signal for a limited set of frequency bands. Transmitting these envelopes only requires several samples per second, whereas the original audio requires thousands of samples per second. On the receiving end, these envelopes, combined with a specific carrier signal, can be reconstructed into intelligible speech. Because the vocoder can be precisely controlled and parameterized, it is a powerful tool for studying how sound is perceived. This is crucial for understanding how cochlear implants transform acoustic signals and how listeners, especially individuals with normal hearing (NH) in studies, might perceive speech or other sounds as if they were CI users \citep{shannon1995speech}. \\

Cochlear implants are medical devices that restore hearing to individuals with severe to profound deafness and are considered the most successful neuroprosthetic device developed to date \citep{kansaku2021}, with over one million people having been implanted worldwide \citep{nidcd2024}. One of the primary goals of cochlear implants is to partially restore access to speech information, thereby enabling effective communication. While successful, outcomes vary significantly from patient to patient due to factors such as age and the duration, cause, and type of hearing loss. Consequently, response data for studies with CI-users often experience high subject-level variability, which is hard to contain or isolate \citep{blamey2012factors}. Furthermore, clinical trials can only rely on a relatively small population of CI users. This provides a challenging environment in which to evaluate the plethora of design choices available for developing a CI \citep{cychosz2024vocode}. \\

As mentioned, an alternative approach is to present Normal Hearing (NH) listeners with signals processed by a channel vocoder to emulate the signal as if perceived by CI users. In practice, the listener can then attempt to recover the content of the original signal, for example, in comprehension tasks \citep{shannon1995speech}. This general framework has become essential in studying hearing loss and allows us to better understand how individuals with CI perform auditory, speech, and language tasks. Moreover, in addition to providing a much larger patient population for conducting trials, it allows for testing specific experimental conditions in isolation \citep{cychosz2024vocode}. This has enabled researchers to conduct several studies that would have been challenging or impossible to conduct without relying solely on CI users. For example, studies using vocoders have been employed to reveal how degraded speech affects language development \citep{newman2020toddlers} and how deeper cochlear implant (CI) insertion depth enhances speech perception \citep{rosen1999adaptation,shannon1998speech}. Additionally, it provides a way for NH individuals to experience some aspects of the sound quality of Electrical Hearing (EH). However, it should be noted that a vocoder does \emph{not} simulate the experience of wearing a CI. Aside from the social and practical implications of being implanted, there are inherent differences between the healthy acoustic hearing system and EH that the signal processing strategies of a vocoder cannot capture \citep{cychosz2024vocode}. \\

In a parallel branch of CI research, the CI listening experience is studied from a different perspective: simulation with computer models \citep{rattay1986,Frijns1995}. This takes another approach to avoid conducting physical experiments with CI users, which requires considerable effort from the human subjects involved. In addition, digital twins have allowed researchers to model specific effects of the auditory system in response to (electrical) stimulation. For example, 3D models helped uncover the current spread throughout the cochlea when stimulated with a given electrode array \citep{kalkman2022}. Moreover, model studies can provide insight into the human hearing system at the single-fiber level \citep{bruce1999stochastic,rattay2001model}. While inherently an abstraction, models can provide a powerful way to study specific effects, enabling a depth and scale of investigation that is often not possible with animal models—and especially not with live human subjects \citep{3dmodelsreview}.  \\

As previously mentioned, vocoders do not capture all the important biophysical aspects related to perception in CI users, as they are based solely on signal processing techniques. Standard channel vocoders do not consider effects such as single fiber refractoriness, electrode interaction, and electrode-to-neural interface. In addition, vocoder design is often specific to a given implant or speech coding strategy, making evaluating a strategy change problematic \citep{cychosz2024vocode}. \citet{ELBOGHDADY2016} proposed a hybrid between the modelling and vocoder-centric approach. This work used a simple population-based Auditory Nerve Fiber (ANF) model as a preprocessing step to the standard vocoder used by the Advanced Combinatorial Encoder (ACE) strategy. This makes it possible to study the effects of newly developed coding strategies within the same framework as already established methods. While \citet{ELBOGHDADY2016} uses a neural model only as a preprocessing step to a standard channel vocoder, the general methodology follows that of \emph{neural decoding} \citep{JOHNSON2000563}. This approach is analogous to those of \citet{pasley2012reconstructing,akbari2019towards}, which utilize ECoG \citep{penfield1954epilepsy} recordings to reconstruct intelligible speech. Other approaches \citep{park2023, daly2023neural} have used fMRI readings to reconstruct complex musical pieces from brain signals using deep learning techniques. \\  

Besides the fact that no live patients are required to conduct experiments in \emph{model-based} neural decoding, an additional advantage is that there is no need to rely on an imperfect signal, such as those collected via fMRI or ECoG. Simulation with models yields exact spike timings per fiber, allowing for precise measurement of the information transferred to the auditory nerve \citep{johannesen2022}. The recent work by \citet{leclere2023} proposed an information-theorectic framework to asses the information contained in the simulated spiking response of a computational model of the implanted auditory nerve. Their model started from the electrode-neural interface (ENI), i.e., from an electrodogram. It then used optimal reconstruction filters to reconstruct the temporal envelope of amplitude and rate-modulated reference signals from the simulated spike trains, based on the approach by \citet{Warland1997}. \\

In this work, we propose a \emph{general methodology} for decoding the output of neural models into sound. In this sense, we can leverage the advancements of contemporary models \citep{bruce2018,kalkman2022,lyon2011using,denobel2024} to develop a biophysically plausible vocoder that reconstructs sound from neurograms, time-frequency representations of auditory nerve activity \citep{HINES2012306}. Leveraging the relationship between the neurogram and the spectrogram, our method employs an inverse Fourier transformation for reconstruction. This, in principle, allows for \emph{any} neurogram-generating source to be used in the simulation process. We demonstrate this using two different ANF models for normal and electrical hearing, without requiring any ad-hoc parameter tuning. This flexibility also allows for the variation of any parameters in these models to match specific experimental conditions, enabling the evaluation of, for example, new speech coding strategies or implant designs within the same computational framework. Since the input signals are in the same domain as the reconstructed signals, i.e., sound, information-theoretic approaches can be applied to quantify the effects of such developments numerically, which is vital for automated development \citep{back2023evolutionary}. Our main contributions are:
\begin{itemize}
    \item We propose a flexible vocoder framework that reconstructs sound from simulated auditory nerve activity using classic signal processing. 
    \item The framework supports interchangeable auditory models, enabling direct comparison between normal hearing and cochlear implant conditions without requiring model-specific vocoders.
    \item We demonstrate that the vocoder captures characteristic differences between models.
    \item We evaluate perceptual intelligibility using an online Digits-in-Noise (DIN) test and show that our results align with clinical benchmarks.
\end{itemize}

The structure of this paper is as follows. Section \ref{sec:pre} introduces the necessary preliminaries, including relevant background and model components. Section \ref{sec:methods} provides a detailed outline of the proposed framework. Section \ref{sec:exp} presents two experiments that evaluate the reconstructed sound, including a perceptual assessment using the Digits-in-Noise (DIN) test. Finally, Section \ref{sec:conclusion} concludes the paper by discussing the findings and implications.

\section{Preliminaries}
\label{sec:pre}
\subsection{Short-Time Fourier Transform}
\label{sec:stft}
Let $x[t]$ denote a discrete-time signal of length $T$, sampled at a rate of $f_s$ samples per second, where $t = 0, 1, \ldots, T-1$. The Short-Time Fourier Transform (STFT) provides a time-frequency representation of $x[t]$ by analyzing short overlapping segments of the signal, allowing the frequency content to be tracked over time \citep{oppenheim1999discrete}. The STFT is computed by multiplying $x[t]$ with a window function $w[t-t_k]$ centered at time frame $t_k$, treating $x[t] = 0$ for $t$ outside $[0, T-1]$, followed by a Fourier transform. Formally, the STFT and its inverse (ISTFT) are defined as:

\begin{align}
    \text{STFT}(x[t]) &= X[t_k, f_i] = \sum_{t=0}^{T-1} x[t] w[t - t_k] e^{-j 2\pi \frac{f_i}{f_s} t}, \label{eq:stft} \\
    \text{ISTFT}(X) &= \hat{x}[t] \quad \quad = \frac{\sum_k \hat{x}_k[t - t_k] w[t - t_k]}{\sum_k w^2[t - t_k]}, \label{eq:istft}
\end{align}
where $f_i$ represents the physical frequency in Hertz (Hz), $f_s$ is the sampling rate, and $\hat{x}_k[t]$ is the inverse Fourier transform of $X[t_k, f_i]$. When appropriately overlapping windows are used (e.g., satisfying the constant-overlap-add condition \citep{Allen1977}), the ISTFT allows for perfect reconstruction of the original signal $x[t]$. \\

The spectrogram, or more specifically, the \emph{magnitude spectrogram} $S[t_k, f_i]$, is given by the magnitude $|X[t_k, f_i]|$ and represents the amplitude spectral density of the signal. This provides a compact representation that can be used for visualization, feature extraction, and further analysis. \\

\subsection{Griffin-Lim Phase Reconstruction}
\label{sec:griffin}
While the complex-valued STFT $X[t_k, f_i]$ is invertible, the magnitude spectrogram $S[t_k, f_i]$ alone discards phase information, making direct inversion impossible. The Griffin-Lim algorithm \citep{Griffin1984} provides an iterative procedure to estimate the missing phase. Given $S[t_k, f_i]$, Griffin-Lim iteratively refines a complex STFT $\hat{X}[t_k, f_i]$ such that: $$|\hat{X}[t_k, f_i]| \approx S[t_k, f_i]$$ and the inverse STFT of $\hat{X}$ corresponds to the STFT of a valid time-domain signal, i.e.: $$\hat{X} \approx \text{STFT}(\text{ISTFT}(\hat{X})).$$ Once the algorithm converges to a stable solution, or after a predefined number of iterations, the estimated $\hat{X}$ can be used to reconstruct $\hat{x}[t]$. This works because the overlapping windows in the STFT introduce redundancy, causing time-frequency components to share information. In particular, adjacent frames partially overlap in time, and spectral leakage spreads energy across nearby frequencies, allowing the Griffin-Lim algorithm to iteratively estimate phase from the shared structure in the magnitude spectrogram.  

\subsection{Mel Scale}
\label{sec:mel}
The Mel scale, named after the word \emph{melody}, is a perceptual scale of equally spaced pitch intervals \citep{Stevens1937}. It reflects the nonlinear sensitivity of the human auditory system and is typically defined as a quasi-logarithmic function of acoustic frequency. The scale is constructed such that equal distances on the Mel axis correspond to perceptually uniform pitch intervals across a specified frequency range. Since it is derived from perceptual experiments in healthy hearing individuals, several implementations exist; in this work, we use the widely adopted version from \citet{slaney1998matlab}. \\

A Mel spectrogram differs from a standard spectrogram in how it represents frequency. Rather than linearly spaced frequency bins, the frequency axis is distributed according to the Mel scale. A linear-frequency spectrogram can be converted into a Mel spectrogram by applying a Mel filterbank, which projects the spectral content into the Mel frequency domain. This results in finer resolution at lower frequencies and coarser resolution at higher frequencies, yielding a more compact and perceptually meaningful representation, particularly effective in audio and speech processing. 

\subsection{Modelling Electrical Hearing}
\label{sec:ehpre}
\begin{figure}[!t]
\centering
\resizebox{\textwidth}{!}{
    \input{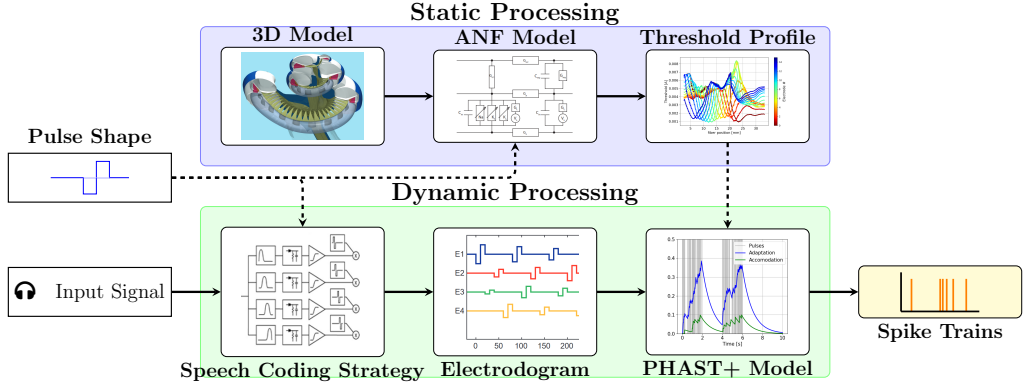}
}
\caption{Diagram of the EH modelling pipeline, consisting of a static and a dynamic part. Solid lines represent data flow, dashed lines denote the exchange of fixed information. The static part of the pipeline calculates a threshold profile for a specific 3D configuration of an implanted cochlea when stimulated with a predefined pulse shape. The dynamic part of the pipeline simulates a temporal response to an incoming input signal, producing spike trains.}
\label{fig:diagram_phast}
\end{figure}

To model EH, we employ a modeling pipeline based on the work of \citet{kalkman2022} and \citet{denobel2024}, utilizing a cascade of biophysical and phenomenological models to generate spike trains for a simulated cochlear implant user. This modeling approach is illustrated schematically in Figure \ref{fig:diagram_phast}, consisting of two main components. 

\paragraph{Static Processing} The first part of the pipeline, indicated with purple in the diagram, models the Electrode-Neuron Interface for a human-implanted cochlea under stimulation with a predefined stimulus waveform (pulse shape). This includes a 3D volume conduction model, which employs a boundary element method to simulate electrical fields in cochleae with arbitrary geometries implanted with multi-channel electrode arrays \citep{Briaire2008Chapter3}. The diagram in Figure \ref{fig:diagram_phast} shows an example of such a geometry. This is followed by a deterministic Auditory Nerve Fiber model \citep{dekker2014, kalkman2022}. This simulates a non-linear double cable model of a human bipolar High Spontaneous Rate fiber \citep{Frijns2000IntegratedUO, briaire2005unraveling} with Schwarz-Reid-Bostock kinetics \citep{Schwarz1995}. This is used to calculate the activation threshold of a fiber when stimulated by a given electrode contact with the predefined pulse shape. Applied for a set of $n_f$ fibers and $n_e$ electrode contacts, this produces a threshold profile, which is a ($n_f \times n_e$) matrix of activation thresholds. 

\paragraph{Dynamic Processing} Where the static part of the pipeline models a fixed stimulation threshold for a single stimulus waveform, the dynamic part simulates a complete temporal response to an incoming audio signal. This includes a Speech Coding Strategy (SCS), configured with the same number of electrode contacts as were used for modeling the 3D geometry,  which generates an electrogram by processing the input signal. The electrodogram, also known as a pulse train, is a multivariate time series comprising $n_e$ channels, where each channel represents the current level of an electrode contact at a specific point in time. This is then used as the input for the PHAST+ model \citep{denobel2024}, a computationally efficient version of the phenomenological model introduced in \citep{vangendt2016fast}. This model converts the pulse trains as generated by an SCS into a simulated spike train, adding temporal behaviour on top of the deterministic thresholds calculated by the ANF model from the static part of the pipeline. The PHAST+ model uses these thresholds to determine the spiking behaviour of an ANF by incorporating the following temporal effects:

\begin{itemize}
    \item \textbf{Accommodation:} A gradual increase in threshold due to sustained stimulation \citep{hodgkin1952quantitative}, modelled by a leaky integrator.
    \item \textbf{Adaptation:} A decrease in firing rate over time in response to prior spiking activity \citep{litvak2001auditory}, modelled as an increase in the threshold by a leaky integrator.
    \item \textbf{Refractoriness:} Temporary inability, or reduced ability of an ANF to fire following a recent spike \citep{yeomans1979}, modeled as an (potentially infinite) increase of the threshold.    
    \item \textbf{Stochasticity:} A stochastic activation threshold \citep{verveen1968fluctuation}. Modelled by a random normal variable with a standard deviation of 5\% of the deterministic threshold, it is used to randomly lower or increase the threshold slightly for each stimulus presentation. 
    \item \textbf{Spontaneous Firing:} Stimulation-independent spontaneous firing behaviour \citep{kiang1965discharge}, modelled by a Poisson process which randomly causes an ANF to produce spikes. This was not included in \citet{denobel2024} and is added to the model specifically for this work. This parameter is set to a constant 50 spikes per second for all modeled fibers. 
\end{itemize}

The code for the PHAST+ model is available as an open-source Python package\footnote{see: \url{https://github.com/jacobdenobel/PHAST}} and includes several pre-processed threshold profiles for different cochlear geometries and electrode arrays. 

\subsubsection{Speech Coding Strategy}
\label{sec:scs}
The speech coding strategy is taken from the Advanced Bionics Generic-Python-Toolbox \citep{jabeimGMT}, modified for interoperability with PHAST+. The code models the Spectral Resolution (SpecRes) strategy, which is a research version of the HiRes Fidelity 120 processing strategy \citep{nogueira2009signal}. The strategy uses asynchronous sequential pulses like Continuous Interleaved Sampling (CIS) \citep{wilson1991better} technique and works via the same fundamental principles:
\begin{itemize}
    \item The incoming signal is divided into several frequency bands using a bandpass filterbank.
    \item Each band’s envelope (the slow-changing amplitude of the signal) is extracted, discarding the fine structure (fast oscillations).
    \item These envelopes are then used to modulate a train of biphasic electrical pulses.
    \item The pulses are delivered sequentially across electrodes, one at a time, in rapid succession. This prevents overlapping stimulation and reduces channel interaction.
\end{itemize}
SpecRes is used to process incoming acoustic signals and generate pulse trains for 16 electrode contacts. Unlike CIS, which assigns one electrode per filter band, SpecRes utilizes current steering \citep{bonham2008current}, which involves the pairwise stimulation of two adjacent electrodes. The strategy uses Fast Fourier Transform (FFT)-based filtering to separate the incoming signal into 15 analysis bands. Due to the limited precision of these filter banks, the strategy effectively acts as a bandpass filter on the acoustic signal, limiting the frequency content from 306 Hz to 8\,054 Hz. Analysis bands span two electrodes, and within each band, a spectral peak locator identifies dominant frequencies. This is used to determine a weighting scheme, where the electrode with its operating frequency closer to that of the estimated peak gets a larger weight.  This enables the creation of so-called virtual electrode channels, which provide higher spectral resolution for the CI user \citep{bonham2008current}. By default, SpecRes separates each analysis band into nine distinct steps between each electrode pair, resulting in a total of 120\footnote{$15 \cdot 9 = 135$ containing $15$ `duplicate' channels} unique virtual channels. For more details, we refer the interested reader to \citet{nogueira2009signal}. 

\subsection{DIN Test}
The Digit-in-Noise (DIN) test \citep{smits2013digits} is a speech-in-noise hearing assessment that measures a listener's ability to recognize spoken digits (typically 0–9) presented against background noise. It is widely used for its simplicity, reliability, and suitability for remote or clinical settings. The test determines the signal-to-noise ratio (SNR) at which a person can correctly identify 50\% of the digit triplets, providing an estimate of speech perception in noisy environments. Because it uses language-independent numerical stimuli, the DIN test is accessible across different populations \citep{polspoel2024global} and has been shown to correlate well with traditional speech-in-noise tests \citep{kwak2021efficacy}. It was originally developed as an online test, and \citet{Shehabi2025} found that the difference between clinical and online testing was not statistically significant for both Arabic and English language speakers.

\section{Methods: The \emph{NeuroVoc} Framework}
\label{sec:methods}
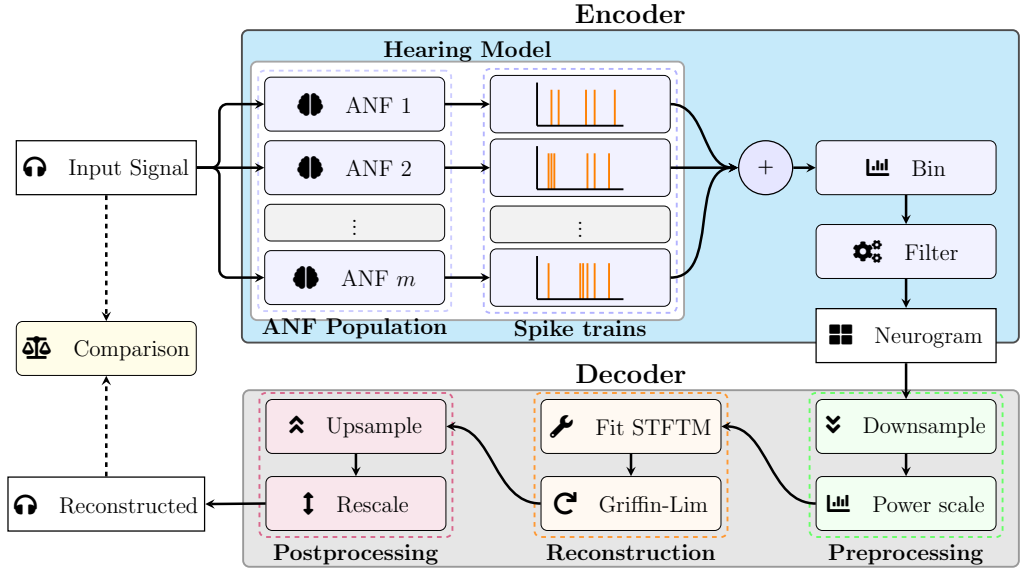
\begin{figure}[!t]
\centering
\resizebox{\textwidth}{!}{
    \begin{tikzpicture}[node distance=1.8cm and 2.2cm]

% Input and comparison nodes
\node (start) [] {};
\node (input) [data, below=of start] {\faHeadphones\quad Input Signal};
\node (compare) [block, below=2.8cm of input, fill=yellow!10] {\faBalanceScale\quad Comparison};

% Population of ANFs block (within encoder)
\node (anf2) [block, right=1.5cm of input, fill=blue!5] {\faBrain\quad ANF 2};
\node (anf1) [block, above=.2cm of anf2, fill=blue!5] {\faBrain\quad ANF 1};
\node (anfdots) [block, below=.2cm of anf2, minimum height=0.8cm, fill=gray!10] {\vdots};
\node (anf3) [block, below=.2cm of anfdots, fill=blue!5] {\faBrain\quad ANF $m$};

\node (spike1) [block, right=1cm of anf1, fill=blue!5] {
\begin{tikzpicture}[x=0.4cm, y=1cm, scale=0.8]
    \foreach \x in {1,1.5,3.4,4,5.4} {
      \draw[very thick, orange] (\x,0) -- (\x,1);
    }
    \draw[very thick] (0,0) -- (6,0);
    \draw[very thick] (0,0) -- (0,1.2);
\end{tikzpicture} 
};
\node (spike2) [block, right=1cmof anf2, fill=blue!5] {
\begin{tikzpicture}[x=0.4cm, y=1cm, scale=0.8]
    \foreach \x in {0.8,1.,1.2,3.5,4,5.} {
      \draw[very thick, orange] (\x,0) -- (\x,1);
    }
    \draw[very thick] (0,0) -- (6,0);
    \draw[very thick] (0,0) -- (0,1.2);
\end{tikzpicture} 
};

\node (spikedots) [block, below=.2cm of spike2, minimum height=0.8cm, fill=gray!10] { \vdots};
\node (spike3) [block, right=1cm of anf3, fill=blue!5] {
\begin{tikzpicture}[x=0.4cm, y=1cm, scale=0.8]
    \foreach \x in {0.8,3.0,3.2,3.5,4,5.} {
      \draw[very thick, orange] (\x,0) -- (\x,1);
    }
    \draw[very thick] (0,0) -- (6,0);
    \draw[very thick] (0,0) -- (0,1.2);
\end{tikzpicture} 
};

% Processing block combining spike trains
\node (combine) [circleplus, right=1.5cm of spike2] {\textbf{+}};
\node (binf) [block, right=.5cm of combine, fill=blue!5] {\faChartBar\quad Bin};
\node (filt) [block, below=.65cm of binf, fill=blue!5] {\faCogs\quad Filter};
\node (neurogram) [data, below=0.65cm of filt] {\faThLarge\quad Neurogram};

% Background for full encoder
\begin{scope}[on background layer]
  \node[draw=black!70, fill=cyan!20, inner ysep=.9cm, inner xsep=.5cm, yshift=.1cm, very thick, rounded corners, fit=(anf1)(anf2)(anf3)(anfdots)(spike1)(spike2)(spike3)(combine)(binf), label=above:{\Large \textbf{Encoder}}] {};
  
  \node[draw=gray!70, fill=white!20, inner sep=.3cm, yshift=0cm, very thick, rounded corners, fit=(anf1)(anf2)(anf3)(anfdots)(spike1)(spike2)(spike3), label={[yshift=-0.4em]above:{\large \textbf{Hearing Model}}}] {};
  
  \node[draw=blue!20, very thick, dashed, rounded corners, fit=(anf1)(anf2)(anf3)(anfdots), label=below:{\large \textbf{ANF Population}}] {};
  \node[draw=blue!30, very thick, dashed, rounded corners, fit=(spike1)(spike2)(spike3)(spikedots), label=below:{\large \textbf{Spike trains}}] {};
\end{scope}

% Arrows from input to each ANF model
\draw [arrow] (input.east) -- ++(0.5,0) |- (anf1.west);
\draw [arrow] (input.east) -- (anf2.west);
\draw [arrow] (input.east) -- ++(0.5,0) |- (anf3.west);

% Arrows from each ANF model to its spike train
\draw [arrow] (anf1.east) -- (spike1.west);
\draw [arrow] (anf2.east) -- (spike2.west);
\draw [arrow] (anf3.east) -- (spike3.west);

% Arrows from spike trains to combiner
\draw [arrow] (spike1.east) to[out=0,in=180] (combine.west);
\draw [arrow] (spike2.east) to[out=0,in=180] (combine.west);
\draw [arrow] (spike3.east) to[out=0,in=180] (combine.west);

% Arrow from combiner to neurogram
\draw [arrow] (combine.east) -- (binf.west);
\draw [arrow] (binf.south) -- (filt.north);
\draw [arrow] (filt.south) -- (neurogram.north);

% Decoder - Preprocessing
\node (filter) [block, below=.8cm of neurogram, fill=green!5] {\faAngleDoubleDown\quad Downsample};
\node (spectrum) [block, below=.5cm of filter, fill=green!5] {\faChartBar\quad Power scale};

% Decoder - Reconstruction aligned left of preprocessing
\node (ifft) [block, left=2.1cm of filter, fill=orange!5] {\faWrench\quad Fit STFTM};
\node (griffin) [block, below=.5cm of ifft, fill=orange!5] {\faRedo\quad Griffin-Lim};

% Decoder - Postprocessing aligned left of reconstruction
\node (resample) [block, left=2.1cm of ifft, fill=purple!10] {\faAngleDoubleUp\quad Upsample};
\node (rescale) [block, below=.5cm of resample, fill=purple!10] {\faArrowsAltV\quad Rescale};

% Reconstructed sound aligned with rescale node
\node (recon) [data, below=of compare, yshift=-.44cm] {\faHeadphones\quad Reconstructed};

% Background groups
\begin{scope}[on background layer]
  \node[draw=black!40, fill=gray!20, very thick, inner sep=.5cm, yshift=-.3cm, rounded corners, fit=(filter)(spectrum)(ifft)(griffin)(resample)(rescale), label=above:{\Large \textbf{Decoder}}] {};
  \node[draw=green!60, very thick, dashed, rounded corners, fit=(filter)(spectrum), label=below:{\large \textbf{Preprocessing}}] {};
  \node[draw=orange!80, very thick, dashed, rounded corners, fit=(ifft)(griffin), label=below:{\large \textbf{Reconstruction}}] {};
  \node[draw=purple!60, very thick, dashed, rounded corners, fit=(resample)(rescale), label=below:{\large \textbf{Postprocessing}}] {};
\end{scope}

% Arrows - decoder
\draw [arrow] (neurogram.south) to[out=-90,in=90] (filter.north);
\draw [arrow] (filter.south) -- (spectrum.north);
\draw [arrow] (spectrum.west) to[out=180,in=0] (ifft.east);
\draw [arrow] (ifft.south) -- (griffin.north);
\draw [arrow] (griffin.west) to[out=180,in=0] (resample.east);
\draw [arrow] (resample.south) -- (rescale.north);
\draw [arrow] (rescale.west) -- (recon.east);

% Dashed comparison arrows
\draw [dashedarrow] (input.south) -- (compare.north);
\draw [dashedarrow] (recon.north) -- (compare.south);
\end{tikzpicture}
}
\caption{Schematic of the NeuroVoc architecture. An input sound is encoded by a population of variable auditory nerve fiber (ANF) models, producing spike trains that are binned and filtered to generate a neurogram. The decoder reconstructs the sound through spectral and temporal transformations, enabling comparison with the original signal.}
\label{fig:diagram}
\end{figure}

Figure \ref{fig:diagram} presents the architecture of the proposed biologically inspired neural vocoder, \emph{NeuroVoc}, which follows an encoder–decoder design. The encoder serves as a flexible simulation framework that can incorporate any neural population-based hearing model. The modular design enables substitution of the entire model as well as parametric manipulation, allowing the simulation of diverse experimental conditions such as neural health, implant configuration, and coding strategy. The decoder reconstructs the acoustic signal using only the neurogram as input. The following sections provide a detailed description of each component.

\paragraph{Code availability}
A Python implementation of the method presented in this work, complete with examples, along with all the code necessary to run the experiments and produce the figures included in this paper, is available open-source at \url{https://github.com/jacobdenobel/NeuroVoc}.

\subsection{Encoder: Generating Neural Responses to Sound}
\label{sec:enc}
The encoder simulates peripheral auditory processing using a population of auditory nerve fiber (ANF) models, each characterized by specific parameters, such as characteristic frequency (CF), spontaneous rate, and temporal response profile. These models receive the acoustic input and produce discrete spike trains that reflect the stimulus-driven firing behavior of individual nerve fibers. For this purpose, any model that simulates the peripheral auditory process can be used, as depicted by the white shaded area in Figure \ref{fig:diagram}. Here, both the acoustic model of \citet{bruce2018} (see Section \ref{sec:nh}) and the EH model presented in \citet{denobel2024} (see Section \ref{sec:eh}) are used to demonstrate the principle. \\

While the encoder design is modular, each ANF model must be associated with a defined place along the tonotopic axis — i.e., a mapping to a characteristic frequency. This is essential for constructing the neurogram as a spatio-temporal representation of neural activity. The neurogram, denoted by \n, is a two-dimensional matrix, where each element \ntf captures activity for a specific time-frequency bin, derived from the spike trains of multiple fibers averaged across multiple repetitions. Here, we generate the neurogram from raw spike trains in two steps, binning and filtering. 

\subsubsection{Binning}
\label{sec:binning}
After being presented with a stimulus, each modelled ANF produces a spike train, a sequence of discrete action potentials over time. This is repeated over $k$ repetitions for each of the $m$ ANF models in the population. This produces a total of $mk$ spike trains for each simulation, denoted by $s_i(t)$, where $i \in [0,\dots, mk)$. To generate a time–frequency representation, the spike trains are discretized along both the temporal and frequency dimensions. \\

Temporally, the spike train is divided into fixed-width time bins of size $\Delta t$, yielding spike counts:
\begin{equation}
    b_i[t_k] = \int_{t_k\Delta t}^{(t_k + 1)\Delta t} s_i(t) dt,
\end{equation}
where $b_i[t_k] \in \mathbb{N}$ represents the number of spikes for trial $i$ in time bin $t_k$. An additional frequency binning step is performed for trials that share the same frequency bin. For every frequency band $f_i$, the spike counts from each associated trial are pooled:
\begin{equation}
    \n_{f_i}[t_k] = \sum_{i \in \mathcal{T}_{f_i}} b_i[t_k],
\end{equation}
where $\mathcal{T}_{f_i}$ is the set of all trials assigned to frequency band $f_i$. This results in a neurogram, a 2D matrix \ntf, where each element captures the magnitude of neural activity for a given time and frequency band. 

\subsubsection{Filtering}
\label{sec:filtering}
The neurogram \n is smoothed along the time axis using a symmetric Hann window to reduce temporal variability. For each frequency band $f_i$, the smoothed signal is computed as:
\begin{equation}
     \n_{f_i}[t_k] = \frac{1}{\sum_{m} h[m]} \sum_{m} \n_{f_i}[t_k - m] \cdot h[m],
\end{equation}
where $h[m]$ is a Hann window of length $m$, given by $h[m] = 0.5 - 0.5 \cos(\frac{2\pi m}{M - 1})$. This operation smooths the neurogram along the temporal axis while preserving its frequency resolution. 

\paragraph{Scaling} The filtered neurogram is normalized by scaling all values \ntf to the range $[0, 1]$ based on the minimum and maximum across the entire matrix, yielding relative activity patterns. 

\subsection{Encoding Neurograms}
\label{sec:encoding}
The encoder was configured with 64 frequency bands, spaced on a Mel scale between 150 Hz and 10\,500 Hz. The same configuration was used wherever possible for both modelling paradigms, i.e., NH and EH. While this is not strictly necessary, it simplifies the configuration and shows generalizability. For each frequency band, ten fibers were simulated, each using 20 independent trials, generating a total of 12,800 spike trains per stimulus condition. Stimuli were presented at 50 dB Root Mean Square (RMS) Sound Pressure Level (SPL)\footnote{This lower presentation level was chosen to avoid the non-linear behaviour the \citet{bruce2018} model shows for louder stimuli. Especially under stimuli with noise conditions, this produces an always-on behaviour for the model (see Figure \ref{fig:choice_noise_ng_bruce}), which severely impacts the reconstruction quality.}, and the binsize of the generated neurograms \n was set to $\Delta t = 36 \mu$s\footnote{The same length as the stimulus waveform used for EH.}. A Hann window of length $H = 1500$, which is $H\cdot \Delta t = 0.054$ s\footnote{A multiple of the cycle speed of the SCS, 1500 / 15 = 100 cycles.}, was used for filtering. As mentioned in Section \ref{sec:enc}, to generate \n, each ANF model needs to be associated with a frequency bin $f_i$. This is explained in more detail in Section \ref{sec:nh} for the NH model and in Section \ref{sec:eh} for EH. 

\subsection{Normal Hearing}
\label{sec:nh}
We simulate spike trains under normal hearing (NH) conditions using the auditory nerve fiber (ANF) model developed by \citet{bruce2018}. The model includes an inner hair cell and synapse component that captures realistic peripheral encoding dynamics, including neural adaptation and refractoriness. For each frequency band, two low, two medium, and six high spontaneous rate fibers were simulated. Each fiber’s characteristic frequency (CF) was set to the center frequency of its corresponding frequency band. For other parameters, default values have been used as provided by \citet{BruceModelcode} with the synapse modifications from \citet{bruce2023poster}. \\

\begin{figure}[h]
\centering
\includegraphics[width=\textwidth]{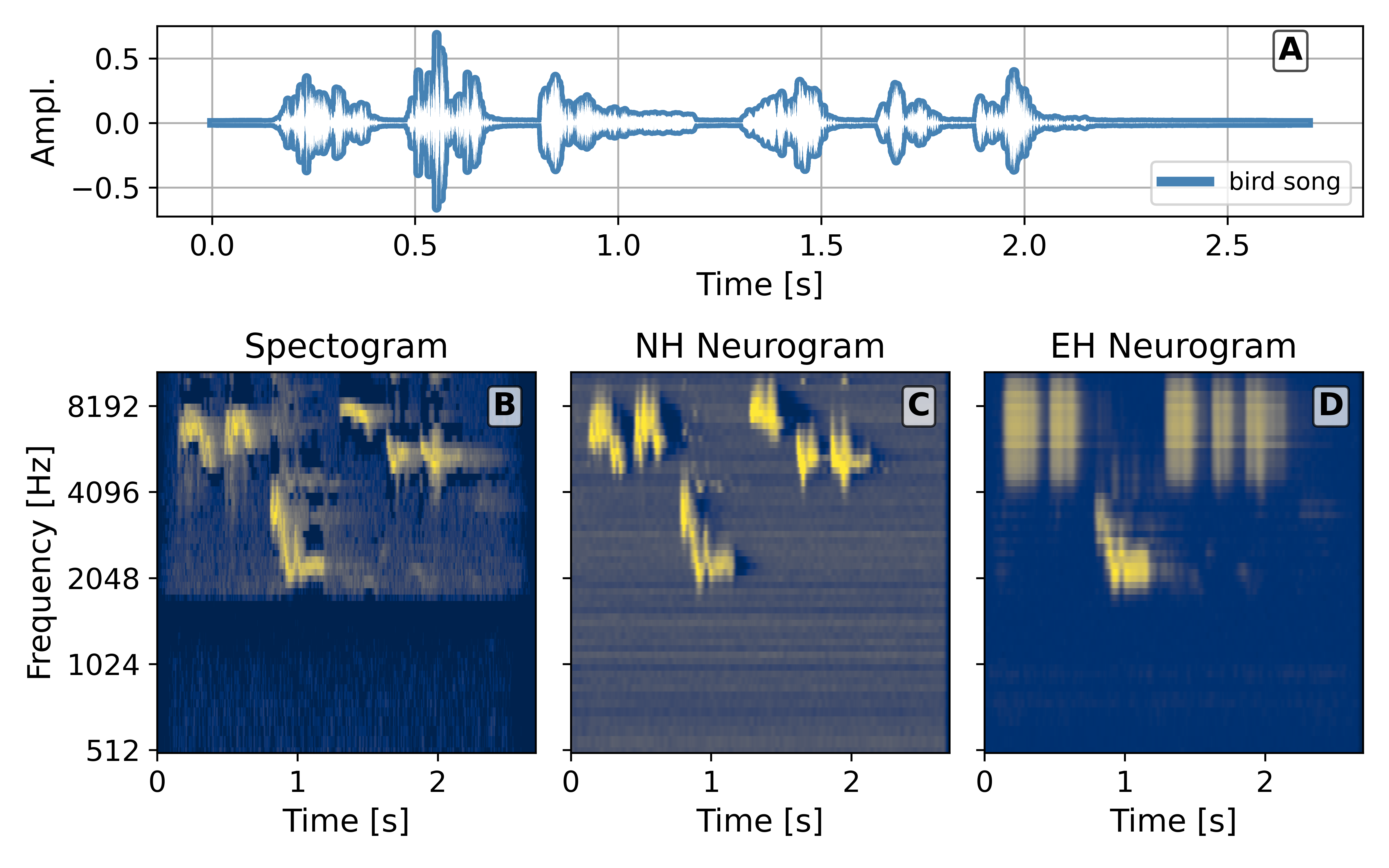}
\caption{Examples of neurograms generated with the two different models. The top figure (A) shows the processed audio stimulus, which is a short fragment of a bird singing. The bottom left figure (B) shows a spectrogram of the stimulus, displayed using a mel scale. C shows the neurogram generated using the \citet{bruce2018} model, Figure D shows a neurogram generated using the EH model described in Section \ref{sec:ehpre}. The color scale of the spectrogram ranges from 0 to -80 dB, and from 0 to 1 for the neurograms. Lighter colors indicate higher values. }
\label{fig:neurogram_example}
\end{figure}

Using the procedure outlined in the previous sections, we generated an example neurogram for a stimulus containing bird song, shown in Figure \ref{fig:neurogram_example}A. From the figure, the relationship between the spectrogram (Figure \ref{fig:neurogram_example}B) and the neurogram (Figure \ref{fig:neurogram_example}C) is clearly visible. Note also that some information is lost, and that the neurogram has sharper transitions than the spectrogram. Additionally, even though the original signal has no noticeable frequency content below 2\,000 Hz, the spontaneous spiking activity does cause the neurogram to have a signal for those frequencies. 

\subsection{Electrical Hearing}
\label{sec:eh} 
\begin{figure}[h]
\centering
\includegraphics[width=0.8\textwidth]{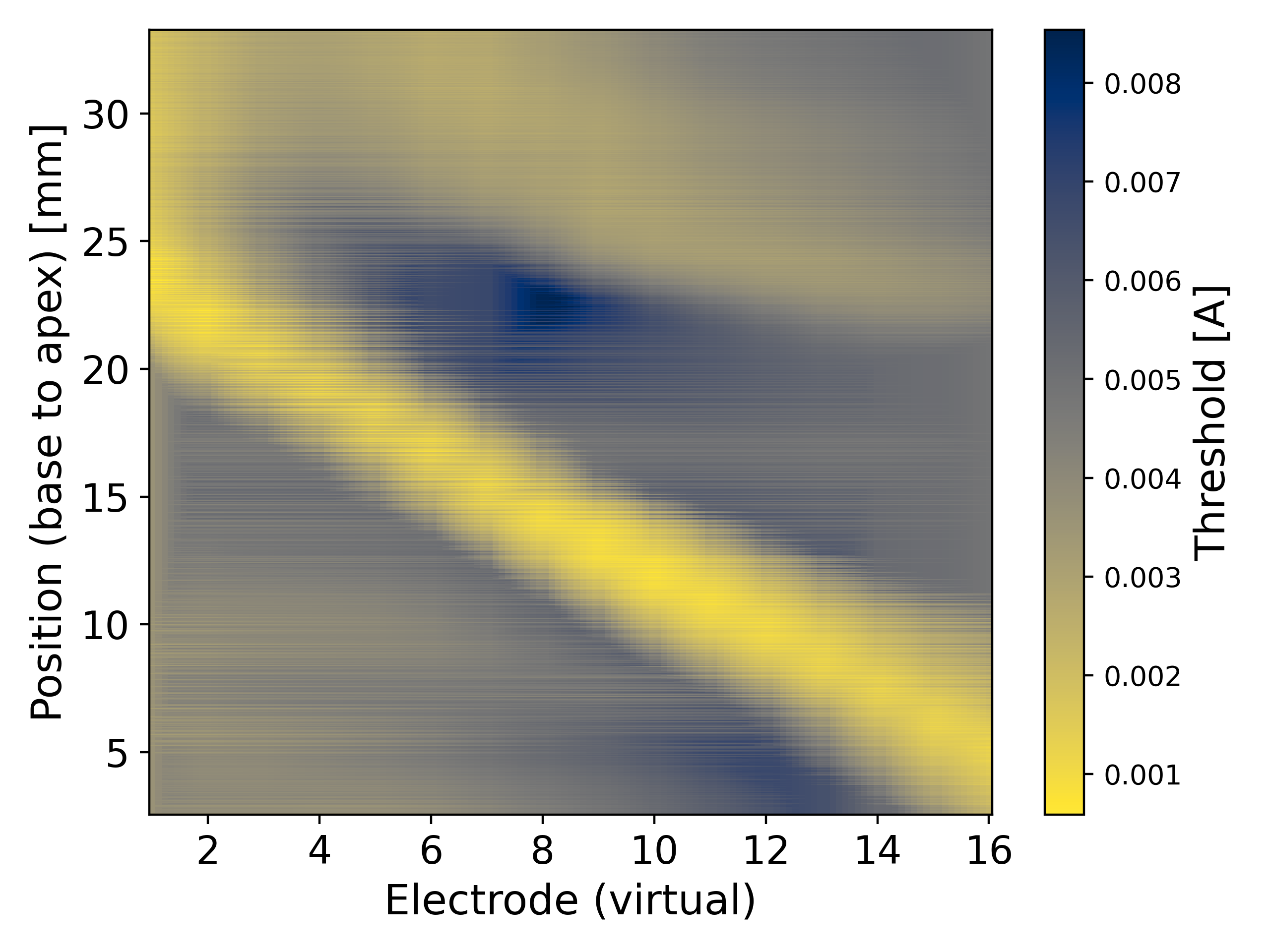}
\caption{Heatmap visualization of a threshold profile specifically generated for SpecRes, containing 135 virtual electrode channels for 3\,200 simulated auditory nerve fibers. The virtual channels are created between two stimulating electrodes, with nine evenly spaced steps. The color indicates the activation threshold of the fiber when stimulated by a given electrode pair. }
\label{fig:thresholds}
\end{figure}

For EH, we use the modeling pipeline presented in Section \ref{sec:ehpre}. The used stimulus waveform is a biphasic cathodic-first square pulse with a phase width of 18 $\mu$s. The modelled geometry (HC3, see: \citet{kalkman2014place}) has a model equivalent of a HiFocus Mid-Scala cochlear implant, which has 16 electrode contacts. The fibers are modeled without neural degeneration \citep{kalkman2022}, and a threshold profile for 3\,200 fibers was generated (see Figure \ref{fig:thresholds}), spaced evenly throughout the cochlea. To accommodate the current steering, 135\footnote{120 + 15 `duplicate' channels} virtual electrode channels were modeled by calculating the activation threshold for a fiber when stimulated simultaneously by two adjacent electrodes. \\

\begin{figure}[h]
\centering
\includegraphics[width=0.8\textwidth]{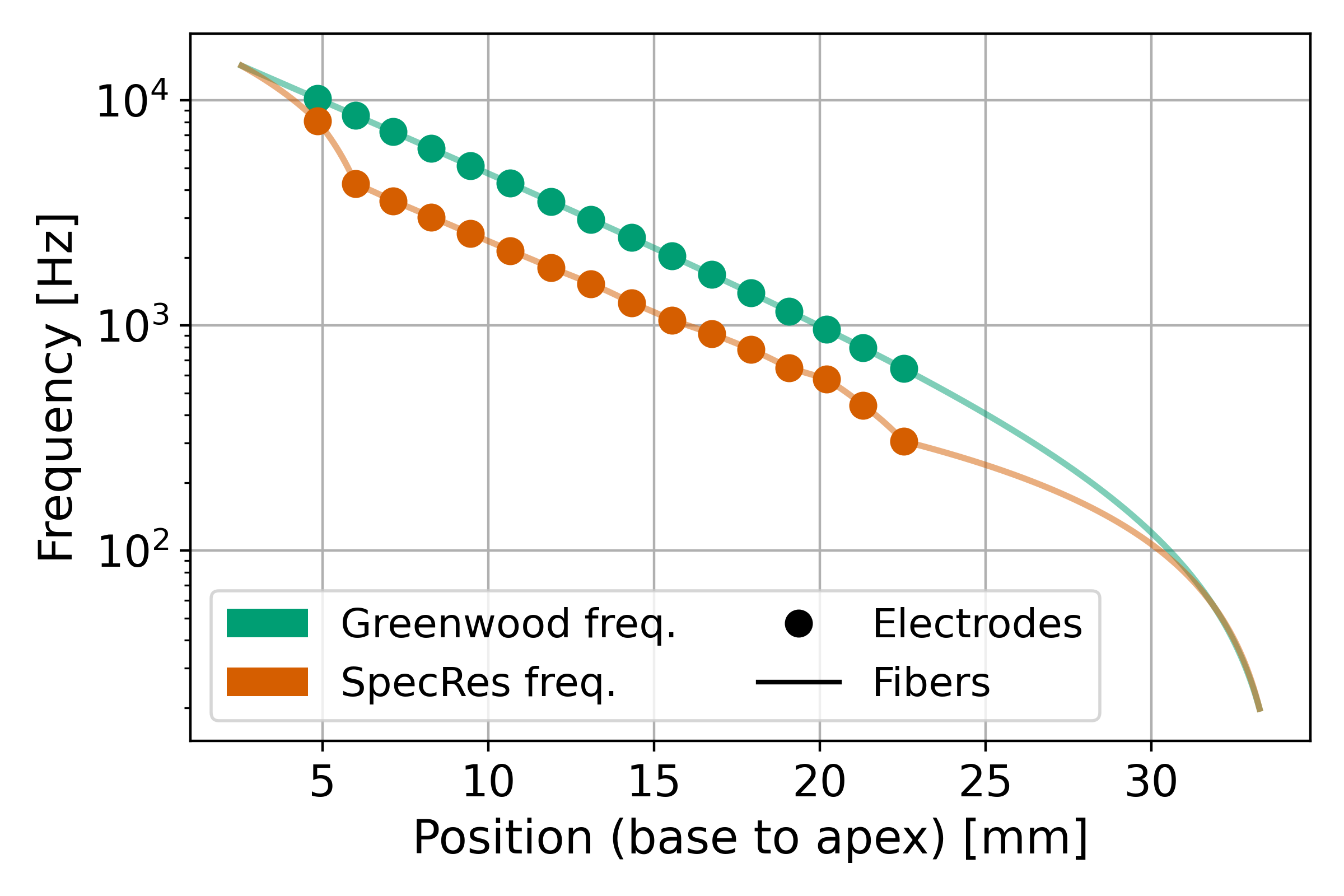}
\caption{Line plot showing the frequency to cochlear position relation for both the stimulating electrodes and the individual fibers. The fibers are shown as a solid line, while the electrodes are visualized using dots. The positional frequency mapping, based on the Greenwood function, is shown in green. The frequency mapping, as used by SpecRes, is shown in orange, with the fiber frequency linearly interpolated to the operating frequencies of the electrodes. }
\label{fig:fiberfreq}
\end{figure}

SpecRes operates at a sampling rate of 17\,400 Hz, which means the Nyquist frequency of 8\,700 Hz effectively cuts off higher frequency content from the audio signal. Additionally, the limited insertion depth of the implant means that low-frequency signals are also not correctly transferred to the CI user. Moreover, each electrode carries the band-pass filtered signal of a specific frequency band, which does not necessarily correspond to the tonotopic location of the electrode. This is illustrated in Figure \ref{fig:fiberfreq}, which shows the mismatch between the signal transmitted by each electrode contact and the tonotopic organization of the cochlea, as predicted by the Greenwood function. From the figure, it can be seen that the signal transmitted by the implant is generally around one octave lower than the tonotopic frequency of the neurons stimulated by that contact \citep{carlyon2010pitch}. This is one of the reasons that a CI can sound too high-pitched, especially for new users \citep{mertens2022smaller}. However, over time, neural adaptation can allow the brain to adapt to a new tonotopic map and reassign meaning to the frequencies, thereby normalizing pitch perception \citep{reiss2007changes}. While this is not the case for all CI users, we take this as a given in assigning a frequency to a fiber. Specifically, we remap the original Greenwood frequencies to the electrode-specific operating frequencies used by SpecRes. This is what is shown by the orange line in Figure \ref{fig:fiberfreq}. We use interpolation to create a continuous frequency profile for all modeled fibers, smoothly transitioning from the natural Greenwood frequencies to the `learned' frequency-place assignments dictated by the implant’s electrode configuration. \\

Based on this `learned' frequency mapping, we randomly select ten fibers for each frequency band $f_i$ that have a frequency mapping falling within that band. We simulate with the parameters of PHAST+ as specified by the `Average Fiber' in \citet{denobel2024}. \\

Figure \ref{fig:neurogram_example}D shows the neurogram generated by the EH model in response to the same stimulus containing bird song. From the figure, it is clear that while the frequency alignment of the model is appropriate, due to the `learned` frequencies, it has a much lower temporal precision than the NH neurogram. This is partly due to CIS, which requires that all electrode pairs be stimulated for a single cycle of the strategy. Moreover, since there is only one electrode pair that stimulates signals over 4\,248 Hz (see Table \ref{tab:analysisbands} in the Appendix), there is very little precision for high-frequency stimuli.

\subsection{Decoding Neural Responses}
The encoder evaluates an arbitrary simulation model and generates a binned and filtered neurogram \n. The decoder component of NeuroVoc (see Figure \ref{fig:diagram}) then performs three sequential operations to reconstruct a time-domain signal $\hat x[t]$. These operations include preprocessing, reconstruction, and postprocessing, and will be explained in more detail below. 

\subsubsection{Preprocessing}
\label{sec:preprocessing}
To reduce the computational cost of the reconstruction stage, the neurogram is first downsampled along the temporal axis using polyphase filtering. If we have the original neurogram \n, which consists of a total of $N$ time frames, we first compute the target number of time frames $n_s = \left\lceil \frac{N}{32} \right\rceil$. Here, 32 represents a fixed hop size, which is the number of frames to skip\footnote{The same hop size that is used in the reconstruction stage}. Resampling was performed with a rational factor $\frac{n_s}{N}$, reduced to its lowest terms, and included an anti-aliasing low-pass filter to minimize spectral distortion. \\ 

After resampling, each neurogram value \ntf was clipped to ensure all values remained within $[0, 1]$, and rescaled to a decibel-like range using a linear mapping: 
\begin{equation}
    \ntf = -80 + 80 \times \min\left(1, \max\left(0, \ntf\right)\right),
\end{equation}
preserving the relative magnitudes. A floor of -80 dB relative to full scale (0 dB) is imposed to suppress irrelevant low-energy content. This threshold corresponds to the default dynamic range in standard signal processing toolkits \citep{librosa2015}. Finally, the decibel-scaled values were converted to a power scale relative to a 50 dB reference, according to:
\begin{equation}
    \ntf = 50.0 \times 10^{\frac{1}{10}\ntf}
\end{equation}
This transformation yields a representation analogous to a Mel-band spectrogram, i.e., power in Mel bands over time, serving as the input to the reconstruction stage. 

\subsubsection{Reconstruction}
\label{sec:reconstruction}
The goal of the reconstruction stage is to recover a time-domain waveform from the processed neurogram representation. Given a rescaled and downsampled neurogram \n, the first step in this process is to retrieve a magnitude spectrogram with a linear frequency scale (see Section \ref{sec:stft}). Currently, the frequency bins of \n are on a Mel scale, and for the signal reconstruction stage, we require it to use the same scaling as an STFT. \\

To accomplish this, we construct a Mel filterbank $\mathcal{M}$, which defines a linear transformation, i.e., a 2D matrix, from FFT bins to Mel-frequency bins. The filterbank maps 512-point FFTs to the frequency scale used in the encoder. To estimate the underlying STFT power spectrum, we solve a non-negative least squares (NNLS) problem:
$$
\argmin_{\hat{S^2} \geq 0} \|\mathcal{M} \hat{S^2} - \widetilde{N}\|_F,
$$
where $\hat{S^2}$ denotes the estimated power spectrum and $\| \cdot \|_F$ the Frobenius norm. The resulting estimate is then converted to a magnitude scaling by taking the elementwise square root. \\

\paragraph{Griffin Lim} After estimating the STFT amplitude spectrum $\hat{S}$, the final step is to reconstruct a time-domain waveform $\hat x[t]$. Since phase information is not available in the generated magnitude spectrum, phase reconstruction is performed using the Griffin-Lim algorithm, as described in Section~\ref{sec:griffin}. Here, the version proposed by \citet{Perraudin2013} was used, and the algorithm was configured with a 512-point Hann window, executed for 320 iterations. A small hop size of 32 samples was chosen relative to the 512-sample window. This increases the redundancy between each consecutive frame and enhances the stability of the algorithm. The hop size matches the downsampling factor applied earlier to the neurogram, ensuring that the reconstructed waveform $\hat x[t]$ has a sampling rate consistent with the neurogram, namely $1/\Delta t$. \\

\subsubsection{Postprocessing}
In the final stage of the pipeline, the reconstructed waveforms $\hat x[t]$ are resampled to the original sampling frequency of the input signal $f_s$. Since the signal is periodic, Fourier-based resampling is used. Finally, $\hat x[t]$ is scaled to 50 dB RMS SPL, producing a reconstructed signal with the same amplitude scaling as the input (see Section \ref{sec:encoding}).

\section{Experiments}
\label{sec:exp}
We perform two experiments to validate whether the proposed approach provides satisfactory reconstructions. First, we examine the spectrograms for a short speech segment and compare the unprocessed sound with the reconstructed sounds using both models (Section \ref{sec:choice}). Secondly, to asses the intelligibility and perceptual quality of the vocoder, we perform an online Digits-in-Noise test, to investigate the synthesized audio from a (normal-hearing) listeners perspective (Sections \ref{sec:dintest} \& \ref{sec:setupdin}).

\subsection{`Choice'}
\label{sec:choice}
We qualitatively examine the reconstructed signals for the word \emph{choice} in the first experiment. For both the normal hearing (NH) and electrical hearing (EH) models, sounds were reconstructed using the approach outlined in the previous section. \\

\begin{figure}[h]
\centering
\includegraphics[width=\textwidth]{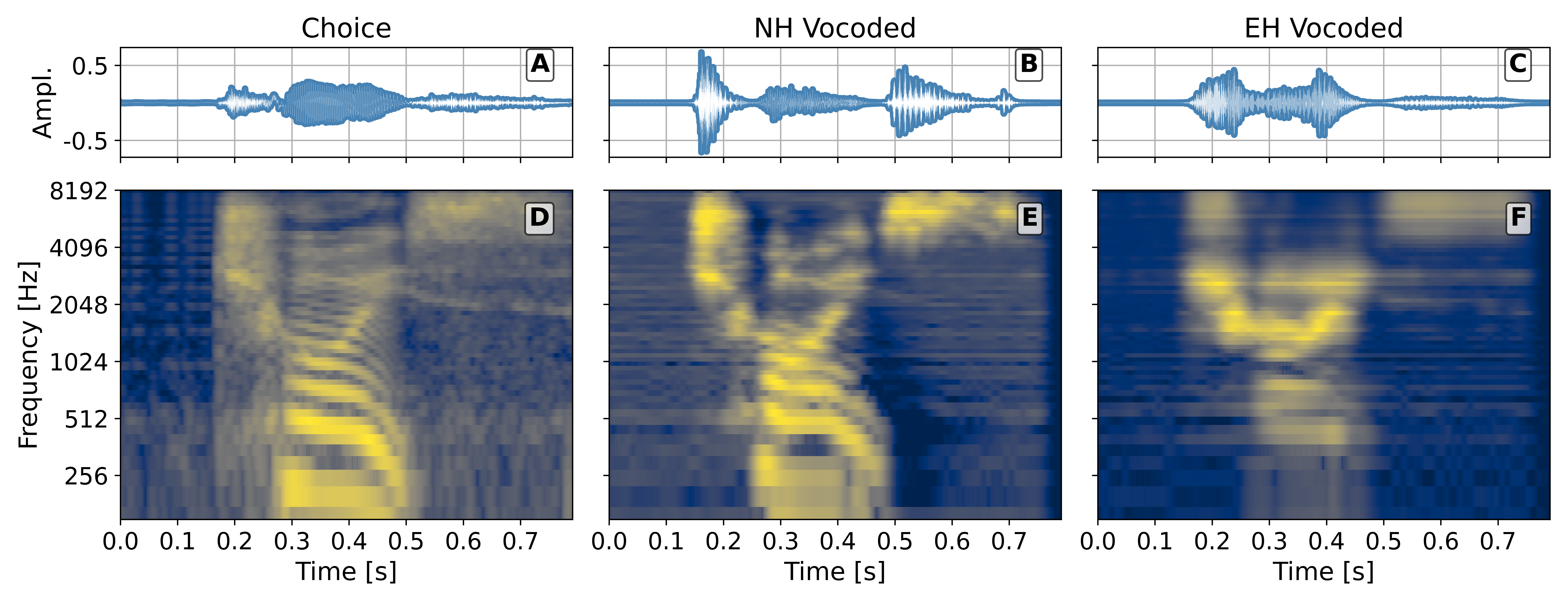}
\caption{Spectrogram and waveform visualizations for the word `choice'. The leftmost (A) panel shows the unprocessed sound. Reconstructed sounds are shown for both the normal hearing (NH) vocoder (B) and the electrical hearing (EH) vocoder (C). Spectrograms, shown in Figures D-E-F, are generated by applying an STFT to the (reconstructed) signal for a 512-point FFT, displaying the magnitude spectrum in a dB scale, ranging from -80 to 0, with lighter colors indicating higher energy. }
\label{fig:choice}
\end{figure}

In the top panel of Figure \ref{fig:choice} (A-C), the reconstructed waveforms are shown alongside the original input stimulus. We observe that while the timing of the reconstructed signals is well aligned with the original, the amplitude is not. Interestingly, the amplitude of the EH reconstruction (Figure \ref{fig:choice}B)  more closely resembles that of the original signal compared to the NH reconstruction (Figure \ref{fig:choice}C). From what we have observed, this is partly due to the response characteristics of the auditory nerve fiber (ANF) model by \citet{bruce2018}, which is sensitive to stimulus onsets following silence. Specifically, even low-intensity inputs can trigger spikes after a period without stimulation. Since perceived loudness in our framework is based on the number of simultaneous spikes within a time-frequency bin, this results in a prominent peak in the reconstructed signal at the onset for NH, even when the input amplitude is relatively low. A similar effect is present in the EH model, but it is less pronounced. \\

When we shift our focus to the (reconstructed) signal's frequency content over time—illustrated in the spectrograms in the bottom row of Figure \ref{fig:choice}—a different picture emerges. Here, the NH vocoder performs remarkably well: it preserves most of the harmonic and spectral content of the input signal, despite the amplitude mismatch observed in the waveform domain. The fundamental frequency and its harmonics are clearly visible and correctly aligned in time and frequency. In contrast, the EH vocoder exhibits substantial spectral degradation. Much of this degradation can be attributed to the limited bandwidth and the coarser frequency binning of the SCS in the CI model. This is especially evident for higher frequencies, transmitted by a single electrode contact, which causes a smearing in the spectrogram. This could also be observed in Figure \ref{fig:neurogram_example}D. In addition, channel interaction, caused by current spread in the cochlea, further degrades frequency selectivity. Because electrical stimulation from one electrode can spread and activate adjacent auditory nerve regions, the effective independence between channels is reduced, leading to overlapping neural excitation patterns and a blurring of spectral details. \\

These differences in reconstruction quality are not unexpected. The NH model is designed to represent the peripheral encoding of sound in a healthy auditory system, while the EH model approximates a CI user. As such, the degraded spectral fidelity observed in the EH reconstructions aligns with each model's intended use: CI users often perceive sound with reduced clarity and resolution compared to normal hearing individuals \citep{bonham2008current}. Therefore, the vocoder results shown here are consistent with the perceptual limitations imposed by the underlying models.

\begin{figure}[h]
\centering
\includegraphics[width=\textwidth]{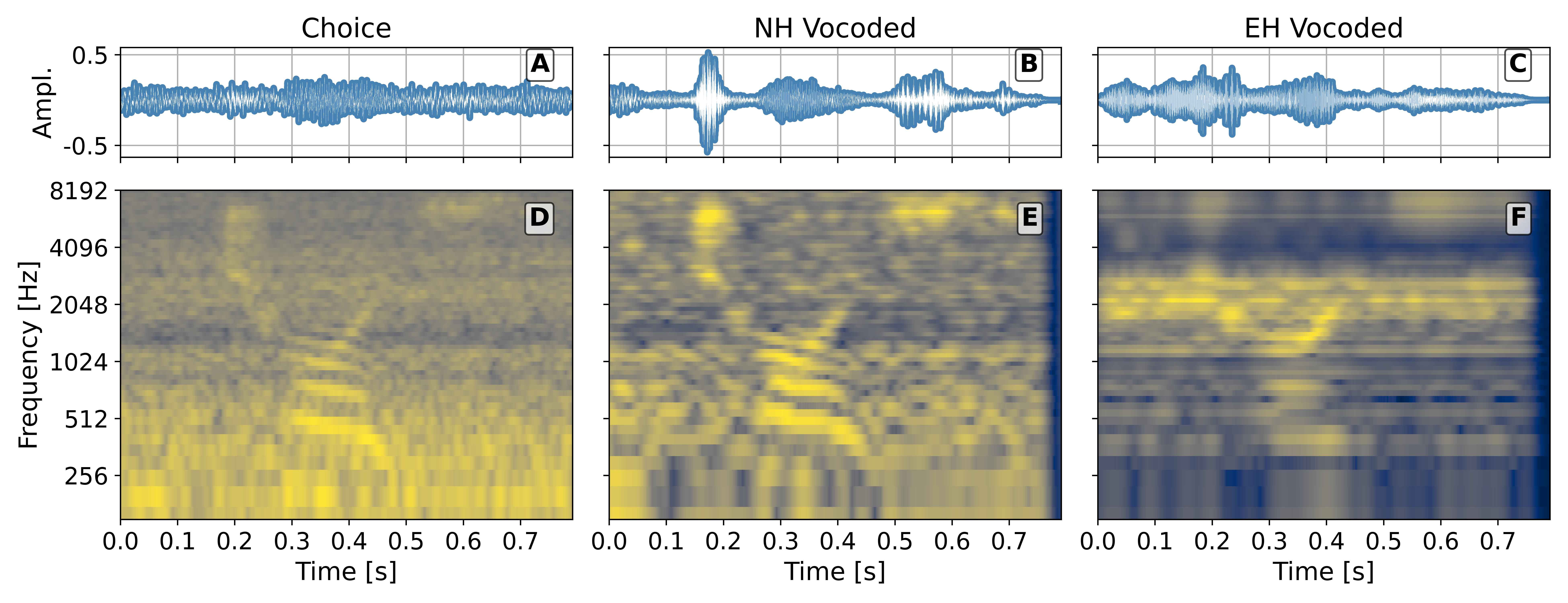}
\caption{Spectrogram and waveform visualizations for the word `choice', mixed with speech-shaped noise at -4 dB SNR. The leftmost panel (A) shows the unprocessed sound. Reconstructed sounds are shown for both the normal hearing (NH) vocoder (B) and the electrical hearing (EH) vocoder (C). Spectrograms, shown in Figures D-E-F, are generated by applying an STFT to the (reconstructed) signal for a 512-point FFT, displaying the magnitude spectrum in a dB scale, ranging from -80 to 0, with lighter colors indicating higher energy. }
\label{fig:choice_noise}
\end{figure}

\paragraph{Adding noise to `choice'} 
When speech-shaped noise is added to the \emph{choice} stimulus at an SNR of –4 dB, the resulting reconstructions are shown in Figure \ref{fig:choice_noise}(A-C). The mismatch in amplitude between the original and reconstructed signals persists for both the NH and EH models. Moreover, we see a clear difference if we compare the amplitude at the beginning of the signal with the amplitude at the end of the reconstructed signal, which both should only contain noise. Specifically, the EH vocoder produces a much larger amplitude at the onset than the (input) signal strength alone would suggest. \\

\begin{figure}[h]
\centering
\includegraphics[width=0.9\textwidth]{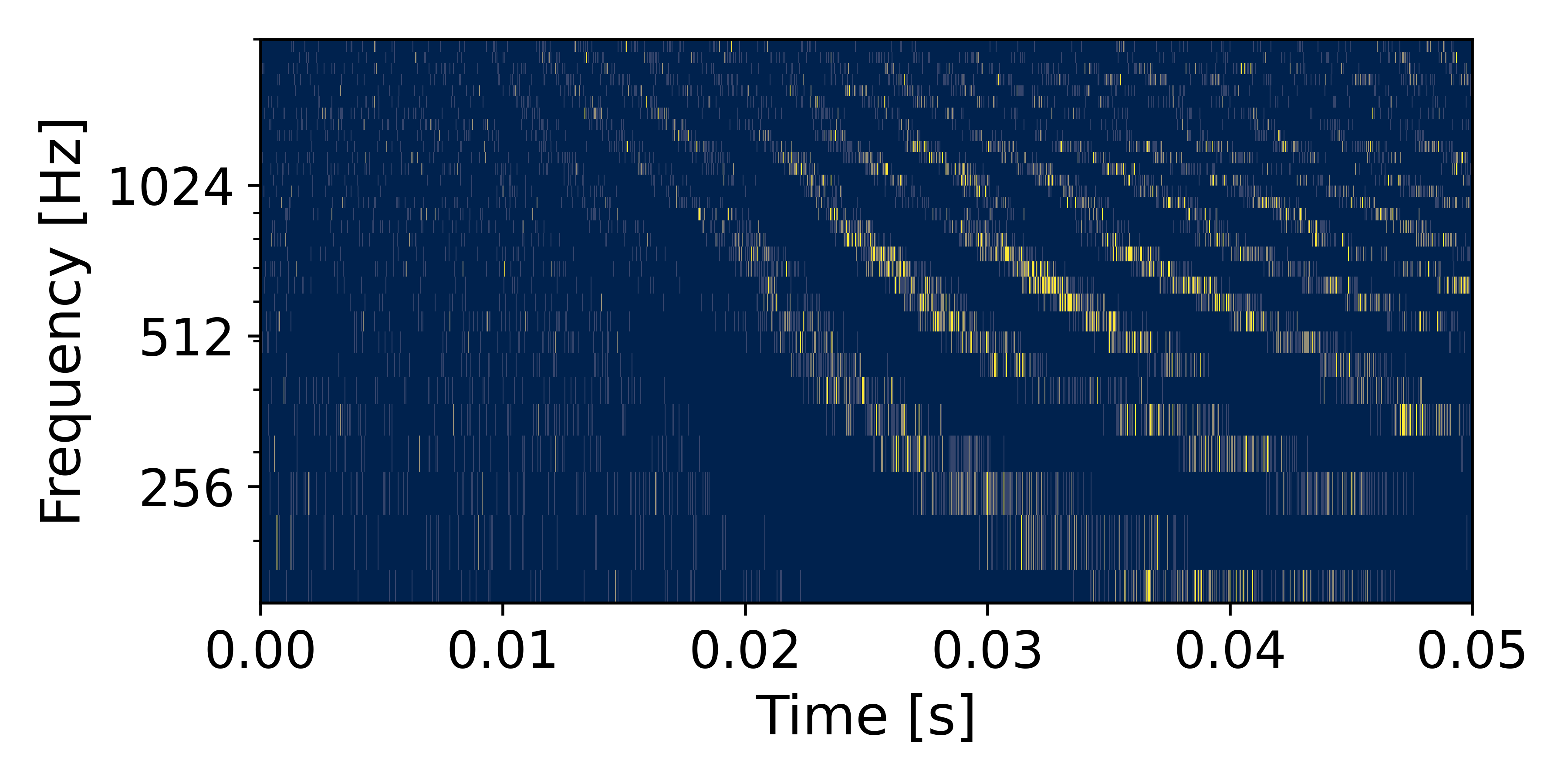}
\caption{Neurogram of the \citet{bruce2018} model for the stimulus `choice', mixed with speech-shaped noise at -4 dB SNR. The first 0.05 seconds of the neurogram are shown, for the fibers with a CF $\in [150, 2000]$ Hz.}
\label{fig:choice_noise_ng_bruce}
\end{figure}

Turning to the spectrograms, several interesting observations can be made. For the NH vocoder, the structure of the original speech signal remains relatively well preserved despite the added noise. However, how the auditory nerve model by \citet{bruce2018} encodes the noise introduces distinct distortions. While the input noise exhibits a relatively flat spectral profile — i.e., consistent energy across frequencies — the reconstructed spectrogram shows irregular ``clumping'' in intensity. Specifically, the energy fluctuates in bursts, alternating between high and low amplitudes over time. This pattern can be attributed to the refractory properties of the auditory nerve fibers. Because each fiber has a recovery period following an action potential, it cannot respond uniformly to a constant or broadband input such as noise. As a result, sustained stimuli like noise are encoded in a temporally modulated way. This behavior is clearly visible in Figure \ref{fig:choice_noise_ng_bruce}, where a zoomed version of the neurogram shows periodic activity interspersed with silent intervals. \\ 

In contrast, this refractory-driven modulation is less evident in the EH vocoder (Figure \ref{fig:choice_noise}F). However, the impact of noise manifests differently: channel interaction becomes substantially more pronounced. Specifically, the spectral smearing in the mid-frequency range (approximately 1\,000–3\,000 Hz) increases, causing certain frequency bands to become overemphasized. This leads to a suppression of finer spectral details and a loss of clarity in the reconstructed signal. \\

Overall, the added noise has a more detrimental effect on the EH model than on the NH model. This is consistent with real-world observations: cochlear implant (CI) users are generally more affected by noisy environments than normal-hearing listeners \citep{cullington2008speech}. The vocoder reconstructions mirror this limitation, reinforcing that the EH model captures key perceptual challenges CI users face.

\subsection{Digits in Noise}
\label{sec:dintest}
This section evaluates two neural vocoders using the Digits-in-Noise (DIN) test \citep{smits2013digits}. The test is based on Dutch speech material consisting of 120 digit triplets. It was conducted online with normal-hearing listeners, each of whom completed three test conditions: the standard DIN test (unprocessed), the test using the NH vocoder, and the test using the EH vocoder. Further details on the test procedure can be found in Section \ref{sec:setupdin}. To begin, we provide an overview of the test data by comparing key statistics of the reconstructed signals to those of the original digit triplets in the next section. 

\subsubsection{Statistics}
\label{sec:dinstats}
In this section, we quantitatively evaluate the neural vocoders on the 120 clean (noiseless) speech stimuli from the Dutch DIN test. To ensure consistent measurements, all audio files—both the original and reconstructed signals—are amplitude-normalized to –20~dB relative to full scale (FS). Before comparison, the reconstructed signals are temporally aligned with their corresponding input stimuli. This is necessary because the neural vocoder introduces a slight delay: ANFs respond only after a stimulus occurs, causing a small timing offset. We address this by applying Dynamic Time Warping (DTW) \citep{berndt1994dtw}, which non-linearly aligns each reconstructed signal with its original. After alignment, we evaluate the reconstructions using two objective measures:

\begin{enumerate}
    \item \textbf{Mean Square Error (MSE)} between the input waveform $x[t]$ and the reconstructed waveform $\hat{x}[t]$, defined as:
    \begin{equation}
        \text{MSE} = \frac{1}{T} \sum_{t=1}^{T} \left(x[t] - \hat{x}[t]\right)^2
    \end{equation}
    This metric quantifies amplitude deviations and reflects how well the reconstructed waveform preserves the dynamic range of the original signal.

    \item \textbf{Mel-Cepstral Distortion (MCD)}, which compares the mel-cepstral coefficient (MCC) sequences of the original and reconstructed signals. MCC sequences represent the spectral envelope of a sound signal. They are calculated by applying a discrete cosine transform to the log-scaled power spectrum produced by a Fourier transform mapped onto a mel frequency scale. Given $c_t$ and $\hat{c}_t$ as the MCC sequences at frame $t$, for the original and reconstructed signal, respectively, MCD is defined as: 
    \begin{equation}
        \text{MCD} = \frac{10}{\ln 10} \sqrt{2 \sum_{m=1}^{M} \left(c_t^{(m)} - \hat{c}_t^{(m)}\right)^2}
    \end{equation}
    where $M=13$ is the number of mel-cepstral coefficients. MCD is commonly used to assess the quality of parametric speech synthesis systems \citep{kominek2008synthesizer}. A lower MCD indicates that the synthesized mel-cepstral sequence closely matches that of the original speech, suggesting higher perceptual similarity between the synthetic and original signals.
\end{enumerate}

\begin{figure}[h]
\centering
\includegraphics[width=\textwidth]{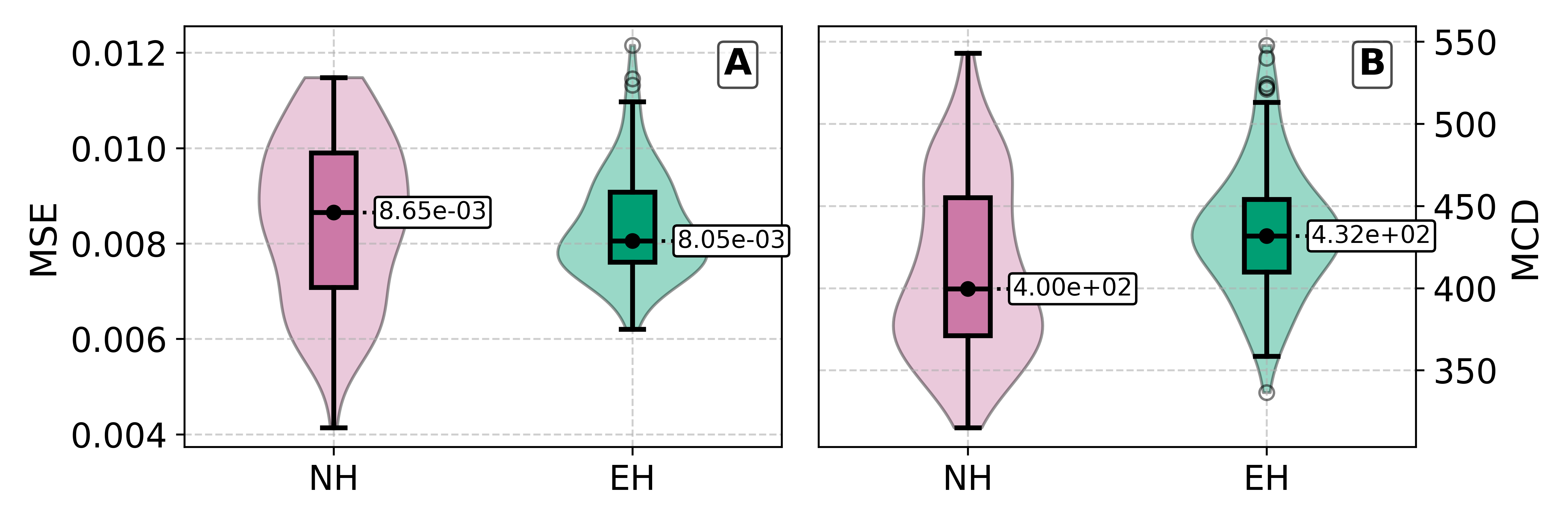}
\caption{Summary statistics of the reconstructed audio samples of the (noiseless) speech material of the digits-in-noise test compared against the original samples. The left panel (A) shows the Mean Square Error (MSE) for both the Normal Hearing (NH) and the Electrical Hearing (EH) vocoder. The right panel (B) shows the Mel-Cepstral Distortion (MCD) for both models. For both, lower is better.}
\label{fig:dinstats_noiseless}
\end{figure}

Together, these metrics provide complementary insights into vocoder performance, similar to the visual analysis presented in the previous section. Specifically, they capture both the temporal amplitude structure via MSE and the preservation of spectral content via MCD. Figure \ref{fig:dinstats_noiseless} presents a boxplot summarizing the results. \\

The patterns observed are consistent with those found in the analysis of the `choice' stimulus (see Section \ref{sec:choice}). For the NH vocoder, the reconstructed waveforms exhibit greater variability in relative amplitude, as reflected by higher MSE values and a larger standard deviation, shown in Figure \ref{fig:dinstats_noiseless}A. In contrast, the EH vocoder shows more stable amplitude reconstruction. \\

However, the opposite trend is observed in the spectral domain, as displayed in Figure \ref{fig:dinstats_noiseless}B. The NH vocoder yields lower mel-cepstral distortion (MCD), indicating superior preservation of the original stimuli's spectral features. By comparison, the EH vocoder shows greater spectral degradation. \\

These differences between the NH and EH vocoders are statistically significant. A two-sided Mann–Whitney U test yielded $p$-values of 0.0197 for MSE and 0.0009 for MCD, confirming that the vocoders differ meaningfully on both temporal and spectral reconstruction metrics for noiseless speech samples, given a confidence bound $\alpha = 0.05$. 

\subsection{Online Digits in Noise Test}
\label{sec:setupdin}
In this section, we describe the experimental setup used to evaluate the intelligibility of vocoder-reconstructed speech using the Digits-in-Noise (DIN) test. A custom web-based testing platform was developed to administer the test following a standardized, adaptive two-up two-down procedure \citep{smits2013digits}. Each participant completed three versions of the test: one using unprocessed stimuli (standard DIN), one using speech reconstructed by the NH vocoder, and one using speech reconstructed by the EH vocoder. The following subsections provide a detailed description of the stimulus preparation, test procedure, and study population.

\subsubsection{Stimulus Preparation}
Each of the 120 Dutch digit triplets used in the Digits-in-Noise (DIN) test was mixed with speech-shaped noise across a range of signal-to-noise ratios (SNRs) to generate the test materials. Every triplet was mixed with a randomly sampled noise instance at SNRs ranging from –20 dB to +10 dB in 2 dB increments, resulting in 16 SNR conditions per digit triplet. This yielded a total of 1,920 noisy speech signals. These signals formed the unprocessed (raw) stimulus set. The same set was then processed through the NH and EH vocoder pipelines, resulting in two additional vocoded stimulus sets —one for each model —yielding three distinct corpora of noisy speech. Mixing and vocoding were performed offline before test deployment, resulting in a consistent body of test stimuli for all users. The resulting stimuli were amplitude-normalized to –20 dB full scale (FS) to control the presentation loudness. 

\subsubsection{Procedure}
The DIN test was implemented on a custom-built website, allowing participants to complete the task remotely using their own devices and headphones. The procedure started with a calibration step adapted from \citet{Shehabi2025}, in which participants adjusted their device volume based on two sentences played 25 dB apart in RMS level: one intended to be clearly audible, and the other loud but not uncomfortable. All subsequent stimuli were presented diotically at a fixed level (–20 dB FS), 5 dB below the high-level sentence and 20 dB above the low-level sentence, ensuring audibility even at the lowest SNR of –20 dB. \\

Following calibration, participants completed a single practice trial to familiarize themselves with the task. The interface was simple and consistent across all test conditions. Each participant completed three DIN tests in randomized order: unprocessed speech (standard DIN), NH-vocoded speech, and EH-vocoded speech. Each test consisted of 24 digit-triplet presentations. For each trial, a stimulus was randomly sampled from the 120-triplet corpus at the current SNR. Each trial began when the participant clicked a single-use playback button, which played the audio. After listening, they selected their response using on-screen digit buttons, with the option to revise their answer before submission. \\

An adaptive two-up two-down procedure was used to vary the SNR based on response accuracy. All tests started at an initial SNR of 0 dB, with values bounded between –20 dB and +10 dB, and were increased by two dB on a correct answer (all digits correct) and decreased otherwise. Performance was quantified using the speech reception threshold (SRT), following the protocol by \citet{smits2013digits}, defined as the average SNR of presentations 5 through 25. The SNR of the 25th presentation is the hypothetical level of the presentation after the last presentation, based on the final adaptive step. \\

All participant data was collected anonymously. Only age, whether participants believed they had normal hearing, and whether they had previously completed the DIN test were recorded. 

\subsubsection{Study population}
A total of 55 participants with self-reported normal hearing completed the study. All participants were fluent in Dutch and completed the test remotely using their own devices and headphones. Data was collected anonymously through the web platform, and no personally identifiable information was recorded. Three participants were excluded from the dataset. Two participants did not complete all three lists, and the other scored an SRT of -5.5 dB on the unprocessed test, which was deemed too high an outlier for NH. 

\subsubsection{Results}
\label{sec:din_results}
The goal of this experiment was to evaluate how well the neural vocoders preserve speech intelligibility (SI) in noise, as measured by the Digits-in-Noise (DIN) test. We hypothesized that: (1) the vocoded conditions would perform in line with expected differences between normal hearing and cochlear implant (CI) listeners; (2) added noise would have a more detrimental effect on the EH (electrical hearing) vocoder compared to the NH (normal hearing) vocoder; and (3) although both vocoders would introduce some degradation, performance in the NH condition would more closely resemble that of the unprocessed (raw) speech condition. \\

\begin{figure}[h]
\centering
\includegraphics[width=\textwidth]{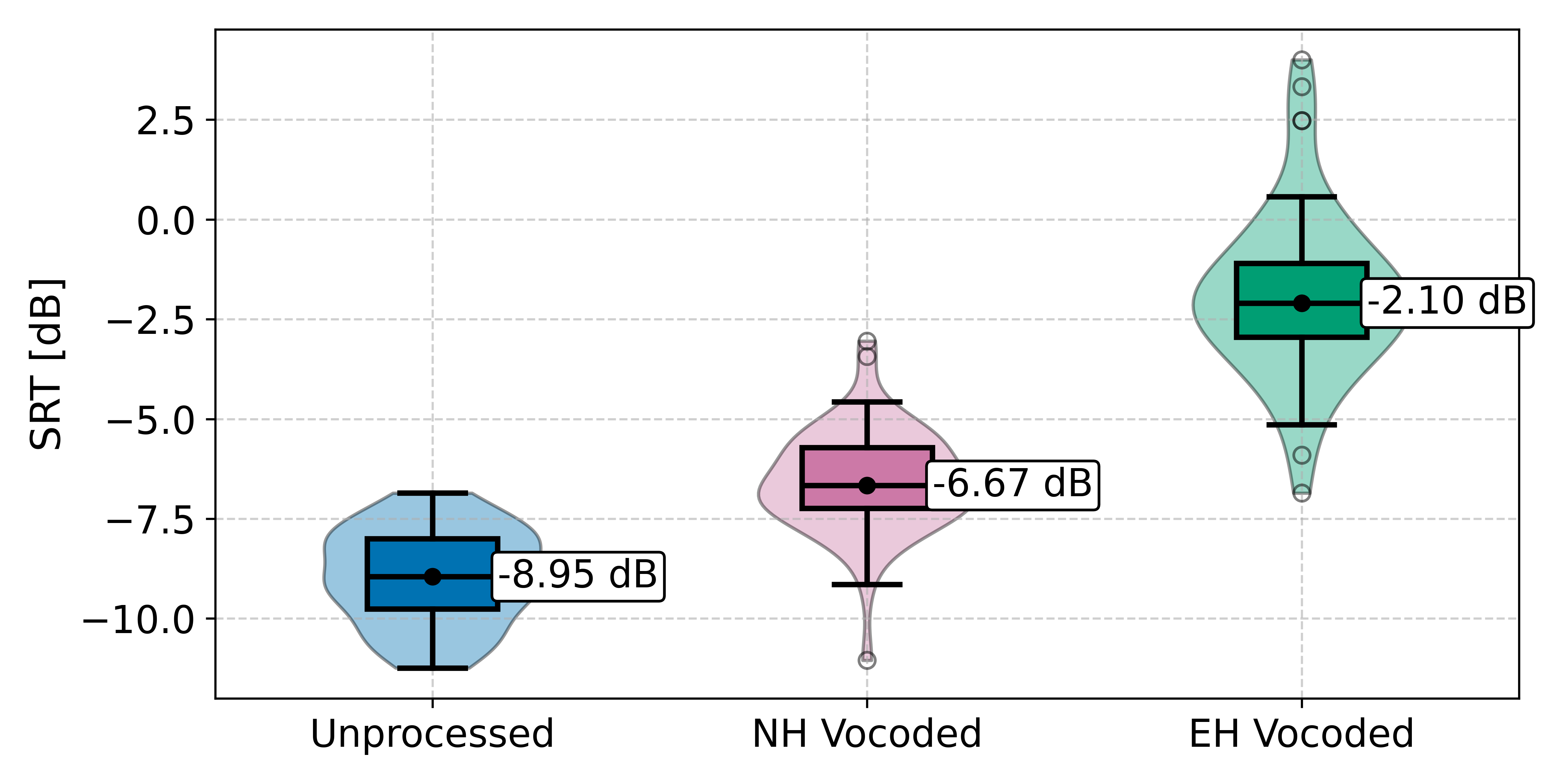}
\caption{Box plot of the speech reception threshold (SRT) for the DIN test. Three test conditions are shown: unprocessed (normal DIN), NH vocoded sound, and EH vocoded sound. The median SRT is provided as an annotation to the plot. }
\label{fig:din_results_boxplot}
\end{figure}

Figure \ref{fig:din_results_boxplot} shows a boxplot of the speech reception thresholds (SRTs) across the three test conditions:  unprocessed, NH vocoded, and EH vocoded. \emph{The results are consistent with our hypotheses}. All groups are statistically different from each other (tested by a Welch’s $t$-test, $\max p \approx 2.5 \cdot 10^{-16}$, $\alpha = 0.05$). The unprocessed condition, i.e., the standard test, yielded the lowest SRTs, indicating the highest intelligibility. The EH vocoded condition significantly elevated the SRT of the participants by 7.12 dB on average. The NH vocoded condition produced intermediate results, with an elevated SRT of 2.4 dB, suggesting better preservation of speech cues in noise but still measurable loss compared to the unprocessed condition. \\

\begin{figure}[h]
\centering
\includegraphics[width=\textwidth]{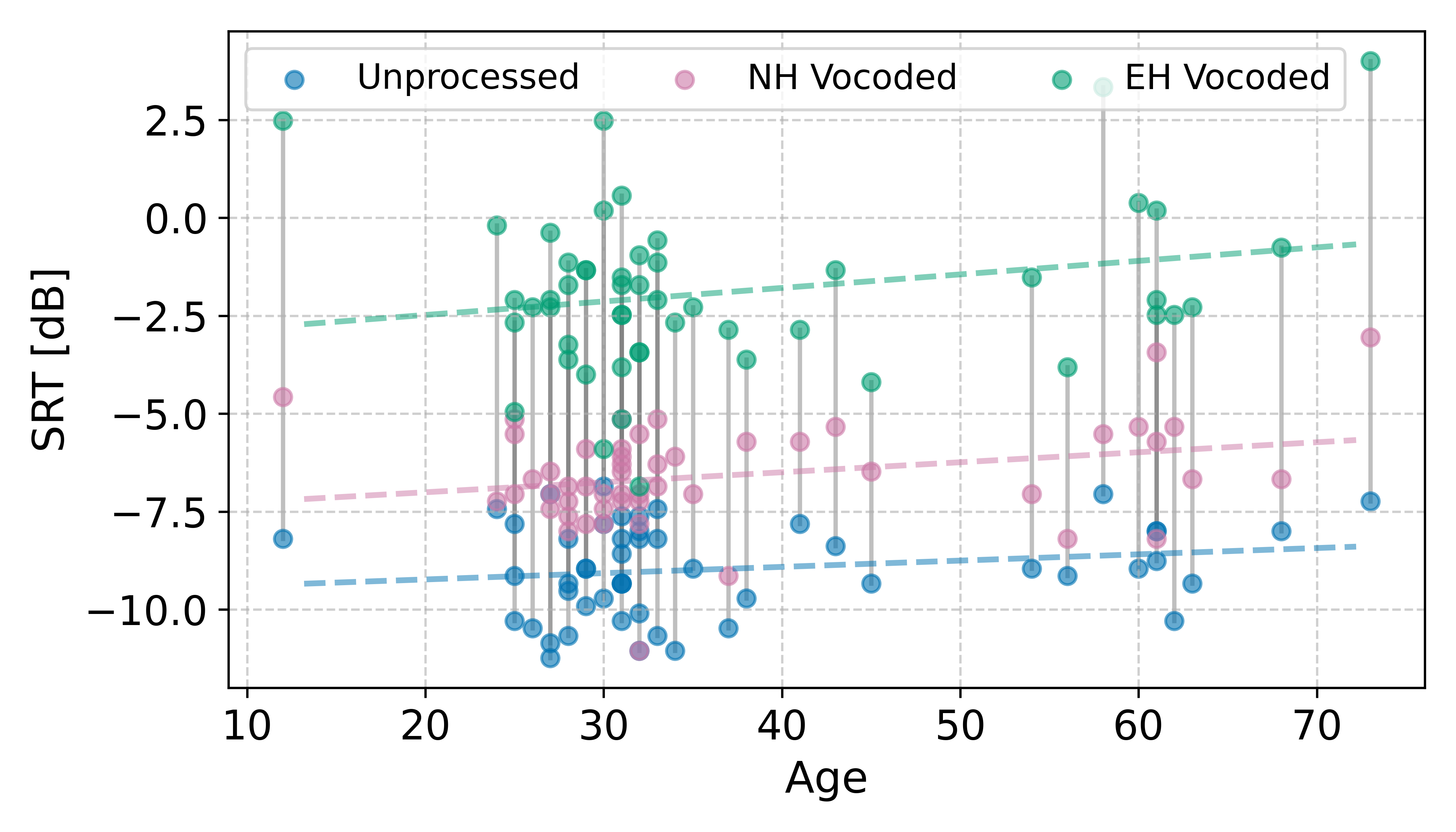}
\caption{Speech reception threshold (SRT) for the DIN test for each of the three test conditions, unprocessed (normal DIN), NH vocoded sound, EH vocoded sound, shown as a function of participant age. A faded grey line connects the test results for each participant across the three conditions. }
\label{fig:din_results_ageplot}
\end{figure}

Figure \ref{fig:din_results_ageplot} shows the effect of age on the test results. There is a minor correlation ($R^2 \approx 0.011$) between age and the DIN test SRT, which aligns with the literature, indicating a negative association between age and SI in noise \citep{GOOSSENS2017109}. Between test conditions, there is only a minor difference in the slope of age vs. SRT, with the slope for the EH vocoded group being the steepest. 

\section{Discussion}
\label{sec:discusion}
In this study, we introduced \emph{NeuroVoc}, a model-agnostic vocoder framework capable of reconstructing acoustic waveforms from simulated neural activity. Unlike recent approaches that rely on machine learning or data-driven methods \citep{park2023, daly2023neural}, our system uses classical signal processing techniques to generate intelligible speech. Its simplicity and modularity allow it to interface with arbitrary ANF models without requiring specialized adaptation. The only requirement of the method is a consistent way of generating a neurogram, which requires a mapping from fiber to frequency. This enables the evaluation of different experimental conditions across diverse modeling paradigms while keeping the vocoder consistent. Such flexibility is particularly valuable for CI research, where specific speech coding strategies often require custom vocoding implementations \citep{cychosz2024vocode}. With our method, these strategies can be evaluated within a unified framework. For example, one could directly compare the SpecRes strategy used in this study to the commonly used ACE strategy, from a different CI manufacturer. Moreover, as demonstrated here, the framework can accommodate entirely different auditory models, such as the normal-hearing model of \citet{bruce2018} and the electrical hearing model present in Section \ref{sec:ehpre}, within the same computational pipeline. Importantly, our method requires minimal parameter tuning. We used default settings for the auditory models and selected vocoder parameters based on generalizability rather than dataset-specific optimization. Despite this, the system performs robustly, indicating that the reconstruction method from neurograms is effective even without fine-tuning. 

\subsection{Reconstruction Quality}
Our results show that the reconstruction quality aligns with the characteristics of the underlying auditory models from which the neurograms were generated. In the normal hearing condition, the reconstructed waveforms were generally of higher quality. Although the amplitude dynamics were somewhat unstable—likely due to pronounced onset responses following periods of silence—the NH vocoder preserved spectral structure well. Harmonic content was clearly represented, resulting in reconstructions with rich frequency detail. Notably, in response to constant stimuli such as noise, the NH model does not produce a steady output due to the refractory behavior of the simulated auditory nerve fibers. This results in temporally ``clumped'' neural activity, which translates into amplitude fluctuations in the reconstructed waveform and reduces the perceived continuity and quality of the sound. \\

The cochlear implant (CI) condition, modeled by the electrical hearing (EH) paradigm, exhibited much reduced temporal and spectral specificity. The limited frequency resolution imposed by the implant and the speech coding strategy used was reflected in the degraded reconstructions. Additionally, spectral smearing, especially under noisy conditions, had a detrimental effect, diminishing signal clarity in affected frequency regions. These results are consistent with known perceptual limitations experienced by CI users, suggesting that the vocoder accurately reproduces the characteristic degradations associated with electrical hearing \citep{shannon1995speech,mertens2022smaller}.   \\

It should also be noted that while many studies \citep{johannesen2022,leclere2023,Gajecki2022} use amplitude-based distance measures such as MSE to evaluate reconstruction quality, this does not necessarily measure intelligibility. For example, as could be observed from Figure \ref{fig:dinstats_noiseless}A, even though the MSE for the EH-vocoder was generally lower than that of the NH-vocoder, the severe spectral distortion imposed by the EH-model caused the reconstructed speech to be much less intelligible, as demonstrated by the results presented in Section \ref{sec:din_results}.  \\

\subsection{Perceptual Evaluation}
Behavioral testing using an online Digits-in-Noise (DIN) test further validated our framework. Participants were presented with unprocessed, NH-vocoded, and EH-vocoded speech stimuli. While each participant only performed a single trial for each of the three test conditions, the learning effect was mitigated over the entire population by randomizing the order of the tests. The relatively large study population strengthens the validity of the results. Participants' SRTs were elevated by approximately 7.1 dB on average for the EH vocoder relative to the unprocessed condition. The NH vocoder condition resulted in only a moderate SRT shift (2.4 dB), indicating that while the vocoder introduces some signal degradation, it still provided for a good reconstruction, even in noisy conditions. 

\subsubsection{Comparing against clinical data}
We observe that our results are similar when compared to published clinical data for the Dutch digit in noise test, as displayed in Table \ref{tab:overview_srt}. There is considerable variability between studies, especially for the groups with a CI. Our results for the unprocessed test are right within the middle of the studies, with a mean value of -8.9 dB. If we pool all the data from the studies together and perform two one-sided Welch’s $t$-tests (TOST), assuming unequal variance, and set the equivalence margin to half the reported SD, i.e. $\frac{0.7}{2} = 0.35$ dB, the results from our unprocessed data are statistically equivalent ($\max p \approx 0.04,\ \alpha=0.05)$. \\
\begin{table}[htbp]
\centering
\caption{Overview of clinical Dutch DIN SRT scores reported in the literature. Data for \citet{Graaff2016Development} was estimated from Fig. 1 (discont. noise, retest). The data for \citet{vroegop2021feasibility} included only children (average age 11.8 $\pm 3.6$), the standard deviation was estimated from Fig. 5. Aggregated values for the mean and standard deviation per group calculated as: $\sum (n_i \cdot \bar x_i) / \sum n_i$  and $\sum ((n_i - 1) \cdot s^2_i) / \sum (n_i - 1)$, where $\bar x_i$ and $s_i^2$ are the reported mean and std. dev. The results from our study have been included for the unprocessed and CI-vocoded groups (see Figure \ref{fig:din_results_boxplot}).}
\label{tab:overview_srt}
\begin{tabular}{l ccc ccc}
\toprule
 & \multicolumn{3}{c}{NH} & \multicolumn{3}{c}{CI} \\
\cmidrule(lr){2-4} \cmidrule(lr){5-7}
Study & $n$ & Mean [dB] & SD & $n$ & Mean [dB] & SD  \\
\midrule
\citet{smits2013digits} & 23 & -8.8  & 0.6    & - & -  & -  \\
\citet{Smits02062016}   & 16 & -9.3  & 0.7 & - & -  & -  \\
\citet{Graaff2016Development} & 12 & -9.5 & 1.0 & 16 & -3.6 & 1.7 \\
\citet{kaandorp2015assessing}  & 12 &  -9.3  & 0.7 & 24 & -1.8 & 2.7  \\
\citet{stronks2025effect} & 18 &  -8.4  & 0.6 & 18 &  -1.5  & 2.5   \\
\citet{vroegop2021feasibility} & - &  -  & - & 58 &  -1.4  & 3.8   \\ \hline
Aggregated & 81 & -9.0 & 0.7 & 116 & -1.8 &  3.1 \\ 
This study & 52 & -8.9 & 1.2 & 52 & -1.9 &  2.1 \\ \hline
\bottomrule
\end{tabular}
\end{table}

Similarly, our results from the EH vocoded test are very close to the average SRT reported in the clinical studies, which are -1.9 dB and -1.8 dB, respectively. Applying the same TOST procedure, using an equivalence margin of $\frac{3.1}{2} = 1.51$ dB, indicates that the data for the EH-vocoded group is statistically equivalent ($\max p \approx 0.0002,\ \alpha=0.05)$ to the clinically reported DIN test scores of CI users. We should note the high subject-level variability within the CI group, which is represented in our test data, with the highest SRT variance found within the EH vocoded group. \\

These findings also reinforce the validity of administering the DIN test online. \citet{Shehabi2025,Graaff2016Development} have shown that DIN results collected remotely under controlled conditions (e.g., with proper calibration and headphone use) do not differ significantly from those obtained in clinical settings. The fact that our unprocessed test aligns with the clinically collected data summarized in Table \ref{tab:overview_srt} suggests that the observed performance differences across vocoder conditions are robust and not an artifact of the online testing environment.  \\

Taken together, these results demonstrate that NeuroVoc provides an effective simulation tool for comparing auditory perception across various hearing conditions.

\subsection{Future work}
While the current study shows that our vocoder method provides realistic reconstructions and captures the characteristics of the used model, several avenues remain open for future work:

\begin{itemize}
    \item Further model-based testing: The framework is well-suited for exploring alternative CI coding strategies. Exploring the impact of different strategies or strategy parameters within the vocoder framework presented here would be a logical next step, and could help easily prototype new methods.

    \item Comparing against standard vocoders: It would be interesting to see how our method compares against vocoders designed explicitly for a given implant/SCS. 

    \item In the current study, the parameters of the method, e.g., number of Mel bands, smoothing filter size, and number of FFT components, are not fully explored. Future work might help uncover more suitable values that are potentially model-specific. 
    
    \item Refractory behavior: The refractoriness in the ANF models limits the ability to encode constant or sustained stimuli. Better spatiotemporal smoothing techniques might help overcome unwarranted temporal modulation in the reconstructed sounds. 
   
    \item Loudness modeling: Our system currently handles relative loudness only, normalized to a suitable range. Incorporating a more accurate loudness scaling based on the firing behavior of the modeled fibers could make the reconstruction more realistic. 
    
\end{itemize}

\section{Conclusion}
\label{sec:conclusion}
Using classic signal processing techniques, we presented a flexible vocoder framework that reconstructs sound from simulated auditory nerve activity. The system supports arbitrary auditory models, enabling direct comparisons between normal hearing and cochlear implant conditions without requiring model-specific vocoders. Our results show that the vocoder captures characteristic differences between models and that reconstructed speech is intelligible, with perceptual performance aligning closely with clinical benchmarks. These findings demonstrate the framework's utility as a lightweight, interpretable tool for prototyping and evaluating auditory perception modelling experiments.

\section*{Acknowledgements}
This collaboration project is co-funded by the PPP Allowance made available by Health~Holland (grant LSHM20101), Top Sector Life Sciences \& Health, to stimulate public-private partnerships. In addition, financial support was also provided by Advanced Bionics Corporation. 
\newpage
\bibliography{main}
\bibliographystyle{apalike}

\newpage
\section*{Appendix} 
\label{sec:appendix}
\begin{table}[h]
\centering
\caption{The 15 analysis bands used by the SpecRes speech coding strategy. The lower and upper bounds of each band are listed in Hz.}
\label{tab:analysisbands}
\begin{tabular}{lll}
\toprule
\textbf{Band} & \textbf{Lower bound} & \textbf{Upper bound} \\
\midrule
1    & 306         & 442         \\
2    & 442         & 578         \\
3    & 578         & 646         \\
4    & 646         & 782         \\
5    & 782         & 918         \\
6    & 918         & 1054        \\
7    & 1054        & 1257        \\
8    & 1257        & 1529        \\
9    & 1529        & 1801        \\
10   & 1801        & 2141        \\
11   & 2141        & 2549        \\
12   & 2549        & 3025        \\
13   & 3025        & 3568        \\
14   & 3568        & 4248        \\
15   & 4248        & 8054       \\ \hline
\bottomrule
\end{tabular}
\end{table}

\end{document}